\documentclass[aps,prb,twocolumn,floats,eqsecnum,showpacs,amsmath,amssymb,floatfix,%
superscriptaddress]{revtex4}
\usepackage[dvips]{color}
\usepackage{graphicx}
\usepackage{epsfig}
\usepackage{dcolumn}
\usepackage{bm}
\usepackage{amsmath}
\usepackage{amssymb}
\begin{document}
\title{Kondo effect in Complex Quantum Dots in the presence of\\
an oscillating and fluctuating gate signal}
\author{M.N. Kiselev}
\affiliation{The Abdus Salam International Centre for Theoretical
Physics, Strada Costiera 11, I-34151 Trieste, Italy}
\author{K. Kikoin}
\affiliation{School of Physics and Astronomy, Tel-Aviv University
69978, Israel}
\author{J. Richert}
\affiliation{Institut de Physique, Universit\'e de Strasbourg, 3,
Rue de l'Universit\'e, 67084 Strasbourg Cedex, France}
\date{\today}

\begin{abstract}
We show how the charge input signal applied to the gate electrode
in a double and triple quantum dot may be converted to a pulse in
the Kondo cotunneling current being a spin response of a nano-device
under a strong Coulomb blockade. The stochastic component of the input
signal results in the infrared cutoff of Kondo transmission.
The stochastization of the orbital component of the Kondo effect in triple
quantum dots results in a noise-induced $SU(4)\to SU(2)$ quantum
transition.
\end{abstract}
\pacs{
  73.23.Hk,
  72.15.Qm,
  73.21.La,
  73.63.-b,
 }
\maketitle

\section{Introduction}
Current interest in charge-spin conversion effects is spurred by
challenging prospects of spintronics. Most of the mechanisms of
such a conversion are related to the interconnection between
electrical and spin current due to spin-orbit interaction,
\cite{Dyak71} which results in spin accumulation near the sample
edges. Such an accumulation in three- and two-dimensional electron
gas in elemental and III-V semiconductors may result in a
spin-Hall effect \cite{Munaz} and positive magnetoresistance.
\cite{Dyak07} It was argued also that the Rashba-type spin-orbit
interaction in a quantum dot assists pure spin current by
modulation of the voltages applied to the leads in a
three-terminal device. \cite{Lug}  Spin Coulomb drag effects
should be mentioned in this context, which may result in spin
polarization of charge current due to intrinsic friction between
electrons with different spin projections induced by Coulomb
scattering. \cite{Cacca76,Davi01}

In all these propositions the possibilities of {\it conversion} of
charge current into spin current were discussed. It is possible
also to try to use the external electric field for the {\it
generation} of spin current or another spin response. One such
idea was formulated recently for light emitting diodes (LED) based
on conjugated polymers, \cite{Li} where dissociation of excitons
in a strong enough electric field may result in the accumulation
of up and down spin densities near the two ends of the LED.

In this paper we show that the charge input signal applied to the
gate electrode in double (DQD) and triple quantum dots (TQD)
forming closed loops (equilateral triangle) may be conversed to a
pulse in Kondo cotunneling current, which is in fact the spin
response of DQD under strong Coulomb blockade. The general idea of
such a conversion was formulated in our previous paper \cite{KKAR}
(hereafter referred as I). Using the example of a T-shaped DQD
occupied by two electrons we have shown that the charge-spin
conversion is possible due to the specific dynamical symmetry of a
spin multiplet consisting of two singlets and a triplet. Since the
states in this multiplet are constrained by Casimir operators of
the group $SO(5)$ characterizing the dynamical
symmetry,\cite{Nova} the time-dependent perturbation in the charge
sector which affects only the singlet states results in a
time-dependent potential acting on the triplet states.

An important aspect of the problem is the interplay between the
coherent (adiabatic) and stochastic components of the input
signal. In paper I we concentrated mainly on the {\it coherent}
signal and discussed in details the conversion of charge noise
into spin response in Kondo tunneling. This response may be
interpreted as dephasing and decoherence of the Kondo screening. Here
we develop the general scheme, where both coherent and stochastic
response of Kondo tunneling to the time-dependent gate voltage are
calculated. Besides, the decoherence due to the noise component of
the input signal is considered in the long relaxation limit
contrary to I, where the white noise approximation was considered
for the noise correlation function.

Another mechanism of charge-to-spin conversion may be realized in
an equilateral triangular TQD in an external magnetic field
penetrating the triangle plane. This model was introduced in Ref.
\onlinecite{KKA} with the purpose of studying the interference
between Kondo tunneling and Aharonov-Bohm interference (see
Ref. \onlinecite{amato08} for an experimental realization). The
actual dynamical symmetry of TQD in a contact with metallic
electrodes is $SU(4)$ because the system possesses two spin and
two orbital degrees of freedom. We show in this paper that the
time dependent gate potential affects only the orbital component of
multiplet, but the Kondo tunneling is sensitive to this
perturbation because both orbital and spin discrete degrees of
freedom are involved in Kondo screening.

\section{Coherent and stochastic charge signal}

\begin{figure}[h]
\includegraphics[width=8.0cm,angle=0]{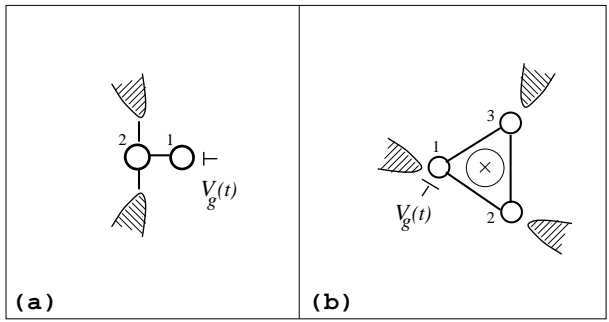}
\caption{(a) Double quantum dot in a T-shape two-terminal geometry
(b) Triangular quantum dot in a three-terminal geometry in a
magnetic field perpendicular to the plane of the triangle.}\label{f.1}
\end{figure}

The subject of our calculations is the study of the transformation of
the charge input signal into a Kondo response of complex quantum dots in
tunnel contact with source and drain electrodes. We study the
mechanism of activation of internal spin degrees of freedom by
means of a time-dependent gate potential applied to passive
electrode. Two examples of complex nano-devices will be
considered. The first is an asymmetric double quantum dot (DQD) which
contacts with metallic leads in a so called T-shape geometry (Fig.
\ref{f.1}a). The second is a triple quantum dot (TQD) in the form of
an equilateral triangle in a three-terminal configuration, where the
bias is established between the dots 2,3 and the time-dependent
gate potential is applied to the dot 1 ((Fig. \ref{f.1}b).

In both cases the electron tunneling is described by the Anderson
Hamiltonian
\begin{equation}\label{1.0}
H= H_{band} + H_{dot} + H_{tun}.
\end{equation}
where three terms are related to band electrons in the leads,
electrons in the complex dot and tunneling coupling between two
subsystems, respectively. The lead Hamiltonian has the form
\begin{equation}\label{hband}
H_{band}=\sum_{b}\varepsilon^{}_{bk}
c^\dag_{bk\sigma}c^{}_{bk\sigma}~,
\end{equation}
Here $b$ stands for the leads, $k,\sigma$ are the wave vector and spin
projection, respectively. The leads are identical in our model,
($\varepsilon_{bk}\equiv \varepsilon_{k}$) so it is convenient to
re-expand the lead electron states $c^\dag_{bk\sigma}$ in terms of
irreducible basis states $c^\dag_{\beta k\sigma}$ of the
corresponding point symmetry groups (mirror symmetry for DQD and
triangular symmetry for TQD).

The Hamiltonian of the dot is written as
\begin{equation}
H_{dot}=\sum_j H_j^0+ \sum^{j\neq j'}_{jj'}H_{jj'}^{0} + H_1(t) ~,
\label{1.1}
\end{equation}
The potential wells in a complex QD are enumerated by the index $j$.
Here
$$
H_j^0=\varepsilon_j n_j +Q_jn_j^2,
$$
describes electron states in the potential well $j$ under the Coulomb
blockade $Q_j$, and $n_j$ are the occupation number operators. The
second term
\begin{equation}\label{hopin}
H_{jj'}^{0}=V\sum_{\sigma}d_{j\sigma}^\dagger d^{}_{j'\sigma}
\end{equation}
stands for the interdot tunneling. By convention $j=1$ is reserved
for the well coupled to the gate. This coupling is described by
the last term in (\ref{1.1}),
$$
 H_1(t)=[V_g(0)+v_g(t)]n_1~,
$$
where the gate voltage contains both a static component $V_g(0)$ and
a time-dependent perturbation $v_g(t)$.

The tunneling term in the Hamiltonian (\ref{1.0}) has the form
\begin{equation}\label{htun}
H_{tun}=W\sum_{j\beta}\sum_{k\sigma}( c^\dagger_{\beta
k\sigma}d^{}_{j\sigma}+ H.c.),
\end{equation}
The values of $j$ and $\beta$ are determined by the geometry
of the complex QD (see below).

We consider a general situation, where the gate potential applied
to a multi-valley complex QD contains both coherent and stochastic
components
\begin{equation}\label{1.3o}
v_g(t) =\tilde v_g(t) + \delta v_g(t).
\end{equation}
The $\tilde v_g(t)$ is the coherent (deterministic) contribution and
$\delta v_g(t)$ is the stochastic noise component which is defined
by its moments
\begin{eqnarray}\label{1.3}
&&\overline{\delta v_g(t)}=0 \\
&&\overline{\delta v_g(t)\delta v_g(t')} = \overline{v^2}f(t-t')
\nonumber
\end{eqnarray}
The overline stands for the ensemble average, the
characteristic function $f(t-t')$ will be specified below.

The starting point of our investigation is the canonical
transformation which converts the gate potential into a
time-dependent operator involving one of the generators of the
group characterizing the dynamical symmetry
of a complex QD.\cite{KA01,KKAR,Nova} This transformation applied
to the Schroedinger operator $-i\hbar(\partial/\partial t)+H_{dot}$
gives\cite{KKAR,KNG,BRUS}
\begin{equation}\label{1.4}
\widetilde{H}_{dot}=U^{-1}_1H_{dot}U_1-i\hbar
U_1^{-1}\frac{\partial U_1}{\partial t},
\end{equation}
with
\begin{equation}\label{u1}
U_1=\exp[i\phi_1(t)n_1]
\end{equation}
and the phase $\phi_1(t)$ given by
\begin{equation}\label{phi1}
\phi_1(t)=\frac{1}{\hbar} \int^tdt^\prime v_g(t^\prime).
\end{equation}
One may apply the Hausdorff expansion to the first term in
(\ref{1.4})
\begin{eqnarray}\label{SW-tr}
\widetilde{H}_{dot}(t)=  H_{dot}^{(0)} + \sum_m \frac{(i)^m}{m!} [
S_1,[S_1...[S_1,H_{dot}^{(0)}]]...],
\end{eqnarray}
where $S_1=\phi_1(t) n_1$, and $H_{dot}^{(0)}$ includes all
time-independent terms from (\ref{1.1})

 It is expedient to introduce "even" and "odd" hopping operators
\begin{equation}\label{thop}
  T_{1j}^{(\pm)}= \sum_{\sigma}[d^\dagger_{1
\sigma}d_{j\sigma}\pm d^\dagger_{j \sigma}d_{1\sigma}]
\end{equation}
To the lowest orders in $V_g$, the time dependent part of the dot
the Hamiltonian acquires the form
\begin{eqnarray}\label{ttun}
\delta H_{\rm
dot}(t)=-V\sum_j\left(i\tilde\phi_1(t){T}_{1j}^{(-)}+
\frac{1}{2}\overline{\phi_1(t)^2}{T}_{1j}^{(+)}\right)
\end{eqnarray}
where
\begin{eqnarray}\label{phi2}
&&\tilde \phi_1(t)=\frac{1}{\hbar} \int^tdt^\prime \tilde
v_g(t^\prime)
\\
&& \overline{\phi_1(t)^2}= \frac{\overline{v^2}}{\hbar^2}\int^t
dt^\prime\int^t
dt^{\prime\prime}f(t^\prime-t^{\prime\prime}).\nonumber
\end{eqnarray}

 Thus, we obtain the effective time-dependent dot
Hamiltonian
\begin{equation}\label{htime}
\widetilde{H}_{\rm dot} = H_{\rm dot}^{(0)}+ \delta H_{\rm coh}(t)
+\delta H_{\rm stoch}(t),
\end{equation}
where the time-dependent perturbation contains a coherent component
[the first term in (\ref{ttun})] and a stochastic one [the
second term in (\ref{ttun})]. In many cases the coherent
perturbation may be easily taken into account in an adiabatic
approximation,\cite{KNG,KKAR} whereas the second term in
(\ref{htime}) results in the stochastization of the quantum state of
the complex QD.

We are interested in the influence of such a time-dependent
perturbation on the electron cotunneling through complex QDs in
Kondo regime. To investigate this influence one should derive the
effective spin Hamiltonian from (\ref{1.0}) by means of a
time-dependent Schrieffer-Wolff (SW) transformation\cite{KNG,KKAR}
taking into account the perturbation $\delta H_{\rm dot}(t)$ in
(\ref{htime}). The adiabatic component of this perturbation
results in temporal oscillations of Kondo transparency and
enhances the tunnel conductance on average,\cite{shut} whereas the
stochastic component of the gate potential is detrimental for
Kondo tunneling. It results in the loss of coherence of the Kondo
singlet state and in the smearing of zero bias anomaly in tunnel
conductance.
 In accordance with the general approach to decoherence
phenomena,\cite{Zeh} one should discriminate between the
decoherence of the ground state of a quantum-mechanical ensemble
and its manifestation at a finite energy/temperature. In the
latter case one should argue in terms of dephasing due to elastic
and inelastic scattering. Both processes are relevant in our
system (see I).

 In the two next sections we will show how the coherent
part of $\delta H(t)$ results in the conversion of the charge
input signal to a Kondo response, and why its stochastic component
brings an end to this process.

\section{Double quantum dot in T-shape geometry. Even electron
occupation}

\begin{figure}[h]
\includegraphics[width=7.5cm,angle=0]{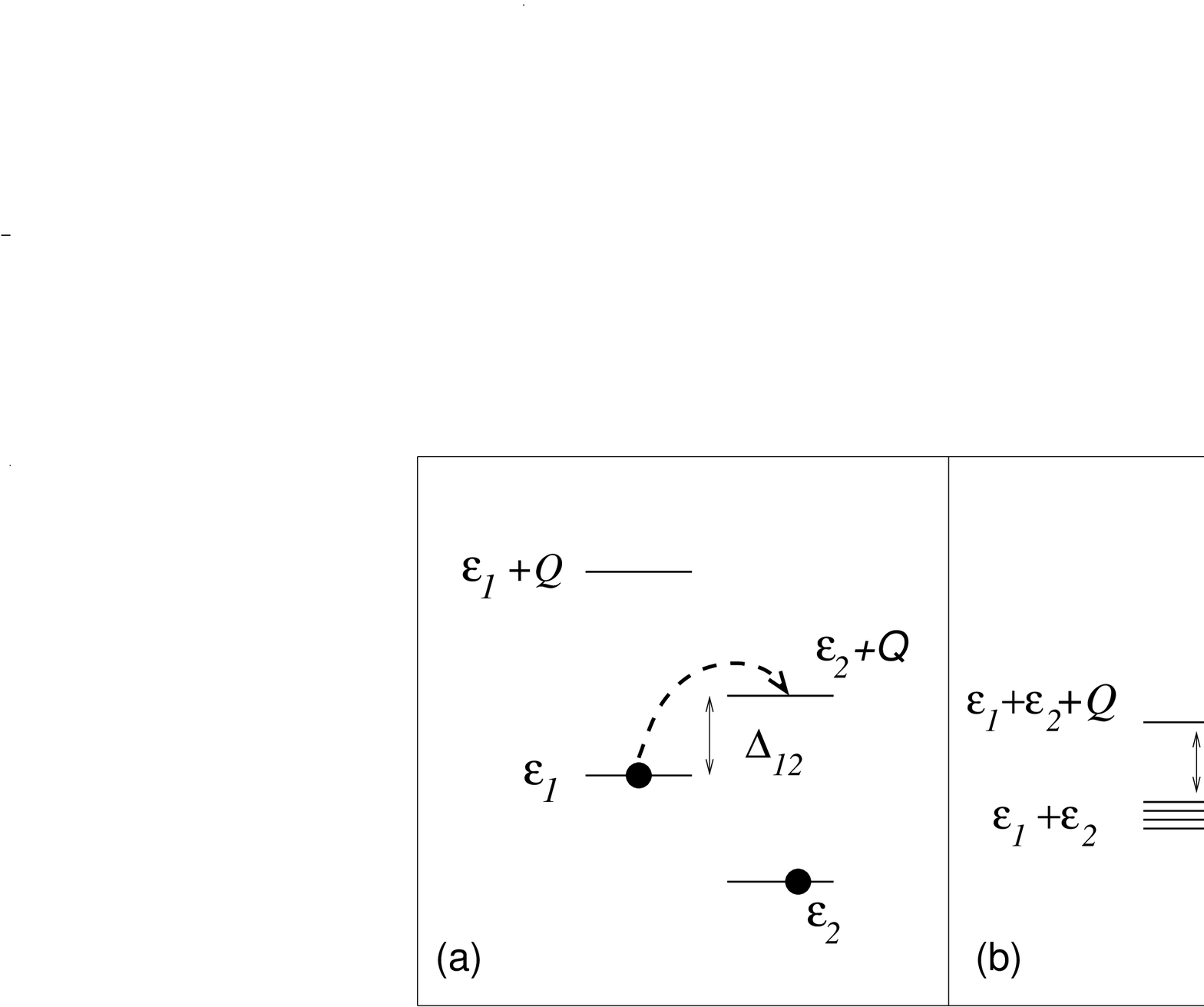}
\caption{Energy levels of an isolated DQD. (a) Single electron
levels. The interdot tunneling $V$ is shown by a dashed arrow; (b)
Two-electron levels $E_\Lambda$; (c) Evolution of $E_\Lambda$ with
the growing of the scaling parameter $\xi=\ln D_0/D$ (see text for
further details).}\label{f.1a}
\end{figure}

We start with the T-shaped DQD (Fig. \ref{f.1}a) and consider the
case of even occupation ${\cal N}=2$ with one electron per
potential well. In this two-terminal geometry the irreducible set
$c^\dag_{\beta k\sigma}$ consists of two combinations of source
($s$) and drain ($d$) leads. Only even standing wave
$c_{ek\sigma}=(c_{sk\sigma }+c_{dk\sigma})/\sqrt{2}$ enters
$H_{\rm tun}$ (\ref{htun}), and we omit in what follows the index
$\beta=e$. The intradot indices $jj'$ have two values 1,2. We
consider a DQD with two equivalent wells ($Q_j=Q$) and assume that
the time-independent component $V_g(0)$ of the gate voltage
modifies the single-electron spectrum in such a way that the
charge transfer excitation due to hopping $H^0_{12}$ (\ref{hopin})
is a relatively soft excitation with the energy $\Delta_{12}\ll Q$
(see Fig. \ref{f.1a}a.)

 Under these conditions the low-lying part
of the energy spectrum of DQD in the charge sector ${\cal N}=2$
consists of two singlet states and one triplet state:\cite{KA01}
\begin{eqnarray}\label{levels}
&&E_S =\varepsilon_1 + \varepsilon_2 -2\alpha V, \nonumber\\
&&E_T =\varepsilon_1 + \varepsilon_2, \\
&&E_E =2\varepsilon_2 +Q_2 + 2\alpha V .\nonumber
\end{eqnarray}
Here $\alpha=V/\Delta_{ES}$ is the effective indirect exchange
parameter, which favors antiparallel orientation of electron spins
in two valleys of the DQD. The ground state of the isolated DQD is the
spin singlet $E_S$.

In spite of this, Kondo tunneling through DQD in T-shape geometry
is possible under certain conditions, because the contact with
lead electrons renormalizes the effective exchange in such a way
that the singlet/triplet level crossing is possible. This result
was obtained in Ref. \onlinecite{KA01} by means of a
renormalization group (RG) technique\cite{Hald} with scaling
parameter $\xi=\ln D_0/D$, where $D_0$ and $D$ is the initial and
current energy scales for electrons in the leads, respectively.

Having in mind this structure of low-lying states, it is
convenient to represent $H_{dot}$ in terms of Hubbard operators
$X^{\Lambda\Lambda'}=|\Lambda\rangle\langle \Lambda'|$, where
$|\Lambda\rangle$ are the eigenvectors corresponding to the
eigenstates $E_\Lambda$ (\ref{levels}):
\begin{eqnarray}\label{hhub}
&&H_{\rm dot}=\sum_{\Lambda} E_\Lambda X^{\Lambda\Lambda}~; \\
&&\Lambda=T\nu, S,E,~~~~\nu=\pm 1,0 \nonumber
\end{eqnarray}
It is important for further calculations that the system of
operators describing transitions between the levels of any
multiplet consisting of two singlets and one triplet forms a
closed set of generators of the $SO(5)$ algebra. Ten generators
forming this algebra are packed into three vectors ${\bf S}$,
${\bf P}$, ${\bf M}$ and one scalar $A$. Vectors describe
transitions within the triplet $T\nu$, between the triplet and the
singlet $S$ and between the triplet and the singlet $E$,
respectively. The scalar $A$ stands for transitions between the
singlets $S$ and $E$. All these operators may be expressed via
Hubbard operators $X^{\Lambda\Lambda'}$ (see Refs.
\onlinecite{KKAR,KKA,Nova} for further details). One may rewrite
the Hamiltonian $H_{\rm dot}$ in terms of these generators
\begin{equation}\label{hdott2}
H_{\rm dot} =  \frac{1}{2}\left(E_T {\bf S}^2 + E_S{\bf P}^2 +
E_E{\bf M}^2 \right) +
 Q({\hat {\cal N}}-2)^2
\end{equation}
Besides, the hopping operator $T_{12}^{(-)}$ (\ref{thop}) may be
represented as
\begin{equation}\label{thop2}
T_{12}^{(-)}=iA\sqrt{2},
\end{equation}
Finally, the Casimir operator $\cal C$ for the $SO(5)$ group is
\begin{equation}\label{constr}
{\cal C} = {\bf S}^2 + {\bf P}^2 + {\bf M}^2 + A^2 =4~.
\end{equation}

Using Eqs. (\ref{hdott2}), (\ref{thop2}) as input data in
(\ref{SW-tr}) and (\ref{ttun}), we obtain the effective
Hamiltonian (\ref{htime}) in the following form (see I)
\begin{equation}\label{hdott22}
H_{\rm dot}(t)= \frac{1}{2}\left(E_T{\bf S}^2 + \tilde E_S(t){\bf
P}^2 + \tilde E_E(t){\bf M}^2 \right) -\mu({\cal C}-4)
\end{equation}
with
\begin{eqnarray}
\label{stadia}
\tilde E_S(t) &=& E_S- \delta_{\rm coh}(t)-\delta_{{\rm stoch},S}(t)  \nonumber \\
\tilde E_E(t)&=& E_E + \delta_{\rm coh}(t)+ \delta_{{\rm
stoch},E}(t).
\end{eqnarray}

Here coherent  and stochastic corrections enter in the form of
time-dependent "energy levels":
\begin{eqnarray}\label{delta}
\delta_{{\rm coh}}(t) =({2V^2}/\Delta_{ES})\phi_1^2(t)~,\nonumber\\
\delta_{{\rm stoch},S}(t)
=({V^2}/4\Delta_{ST})\overline{\phi_1^2(t)}~, \\
\delta_{{\rm stoch},E}(t)
=({V^2}/4\Delta_{ET})\overline{\phi_1^2(t)}~, \nonumber
\end{eqnarray}
 $\Delta_{\Lambda\Lambda'} =|E_\Lambda-E_{\Lambda'}|$.
As was pointed out in Ref. \onlinecite{KKAR}, a charge perturbation
cannot directly affect spin degrees of freedom, and the triplet
level $E_T$ remains time independent.

It is seen from (\ref{stadia}), that the coherent component
results in the time-dependent shift of energy levels. It may be
treated as an adiabatic correction provided the time-dependent
perturbation is weak enough, $\delta_{\rm coh}(t)\ll \Delta_{ET}$.
Below we adopt this adiabatic approximation.

Unlike the coherent renormalization, stochastic components are not
"in phase", i.e., $\delta_{{\rm stoch},S}(t)\neq \delta_{{\rm
stoch},E}(t)$. This inequality makes the constraint
imposed on spin dynamics by the Casimir operator ${\cal C}$
(\ref{constr}) "fragile" for the $SO(5)$ group. Another source of
stochasticity is the last term in Eq. (\ref{1.4}). In lowest
order in the stochastic correlation functions (\ref{1.3}) its
contribution is $\sim \hbar n_1 \overline{\dot\phi_{1}(t)\phi_{1}(t)}$.
This contribution may be converted into additional 4-th order
corrections to Eqs. (\ref{stadia}), and we neglect it in the
following calculations.

A time-dependent SW transformation of the Hamiltonian
(\ref{1.0}) is performed by means of canonical transformation
$\tilde H= U_2HU^{-1}_2$. The phase $\Upsilon$ in the matrix
$U_2=\exp i\Upsilon$ is given by
\begin{equation}\label{2.1}
\Upsilon(t)=\sum_{k\sigma}\left[\upsilon_k^S(t)X^{Sr_{\bar{\sigma}}}
c_{k\sigma}+ \upsilon_k^E(t)X^{Er_{\bar{\sigma}}} c_{k\sigma} -
H.c.\right],
\end{equation}
The coefficients $\upsilon_k^S(t)$, $\upsilon_k^E(t)$ are fixed through
the condition
\begin{equation}\label{1.12}
H_{\rm tun}+ [\Upsilon, (H_{dot}+ H_{band})] =
i\hbar\frac{\partial \Upsilon}{\partial t}.
\end{equation}
The solution of this equation is described in I, and the resulting
cotunneling Hamiltonian has the form
\begin{eqnarray}\label{SW2}
H_{\rm cotun}(t) &=& J^{T}_0 {\bf S}\cdot {\bf s} + J^{S}(t) {\bf
P}\cdot {\bf s}+J^{E}(t) {\bf M}\cdot {\bf s}.
\end{eqnarray}
Again, pure spin scattering is not affected by charge
perturbation, but time-dependent spin-flip transitions in the
leads described by the two last terms in (\ref{SW2}) arise due to
the fact that the dynamical symmetry of the dot spin multiplet is
activated by the time-dependent gate potential. Like in Eq.
$(\ref{hdott22})$, the time-dependent coupling parameters in the
SW Hamiltonian contain both coherent and stochastic components,
\begin{equation}\label{adstoh}
 J^\Lambda(t)=J^\Lambda_0 +J^\Lambda_{\rm ad}(t)+J^\Lambda_{\rm stoch}(t)
 \end{equation}
($\Lambda=S,E$). The time-dependent corrections to $J^\Lambda_0$
are calculated in Appendix A. Now we are well prepared to study
the contribution of adiabatic and stochastic corrections to Kondo
tunneling.

\subsection{Coherent input signal.
}

We study here the transformation of the monochromatic gate
potential
\begin{equation}\label{pergate}
\tilde v_g(t) = \tilde v_g\cos \Omega t.
\end{equation}
into a coherent (adiabatic) Kondo response under the condition
$\Omega\ll \Delta_{ST}$ and $\delta_{\rm coh}(t)\ll \Delta_{ET}$.
Then the phase $\widetilde\phi_1(t)$ has a simple form
\begin{equation}
\widetilde \phi_1(t)=\frac{\widetilde v_g}{\hbar\Omega}\left(\sin
\Omega t-1 \right)
\end{equation}

We calculate the coherent (adiabatic) response at low enough
energy and temperature, where the last term in (\ref{SW2}) may be
omitted. Then we remain in the reduced $\{S,T\}$ part of Hilbert
states, i.e. assume that the Kondo temperature $T_K\sim \Delta_{ST}$
is valid. The dynamical symmetry of the reduced adiabatic (ad) effective
Hamiltonian
\begin{eqnarray}
H_{SW}& = & H_{\rm dot}^{(\rm ad)} + H_{\rm cotun}^{(\rm ad)},\label{SW1} \\
H_{\rm dot}& = & \frac{1}{2}\left(E_T {\bf S}^2 + E_S(t) {\bf P}^2
\right) +
 Q({\hat {\cal N}}-2)^2, \nonumber \\
H_{\rm cotun}^{(\rm ad)} &=& J^{T}_0 {\bf S}\cdot {\bf s} +
[J^S_{0}+ J_{\rm ad}^{S}(t)] {\bf P}\cdot {\bf s} \nonumber
\end{eqnarray}
is $SO(4)$.

The adiabatic part of the time-dependent Hamiltonian may be
incorporated in a Haldane-type RG theory.\cite{KKAR,KNG,Hald} As
a result, the levels $E_\Lambda$ acquire self energies scaled with
the parameter $\xi=\ln (D_0/D)$, namely $M_\Lambda=\alpha_\Lambda
\xi$  so that
\begin{eqnarray}
E_T &\to & E_T -\alpha^{}_T\xi , \label{adiaa}\\
E_S(t)&\to& E_S- \alpha^{}_S(t)\xi , \nonumber
\end{eqnarray}
The self energy $M_S$ depends parametrically on time. The
coefficients $\alpha^{}_{S,T}$ were calculated in Ref.
\onlinecite{KA01}, and it was shown there that the inequality
$\alpha^{}_T
> \alpha^{}_S$ is always valid due to existence of excited singlet level $E_E$.
Due to this inequality, the S/T level crossing may occur at this
stage of renormalization (see Fig. \ref{f.1a}c), so that the ground
state of the system is triplet to the right of the crossing point.
The time dependent factor $\alpha_S(t)$ describes parametrically
slow variations of the scaling trajectory due to adiabatic
(coherent) corrections given by Eqs. (\ref{stadia}) and
(\ref{delta}).

The expression for $J_{\rm ad}^{S}(t)$  derived in Appendix A
reads explicitly (see Eq. \ref{Apex}):
\begin{equation}\label{josc}
J^S_{\rm ad} \approx \frac{\sqrt{2}W^{2}}{\epsilon_2-M_S}
\widetilde{\phi}_1(t)
\end{equation}
(one may neglect the time-dependence of the denominator).

\begin{figure}[h]
\begin{center}\hspace*{-2cm}
\includegraphics[width=12cm,height=9cm,angle=0]{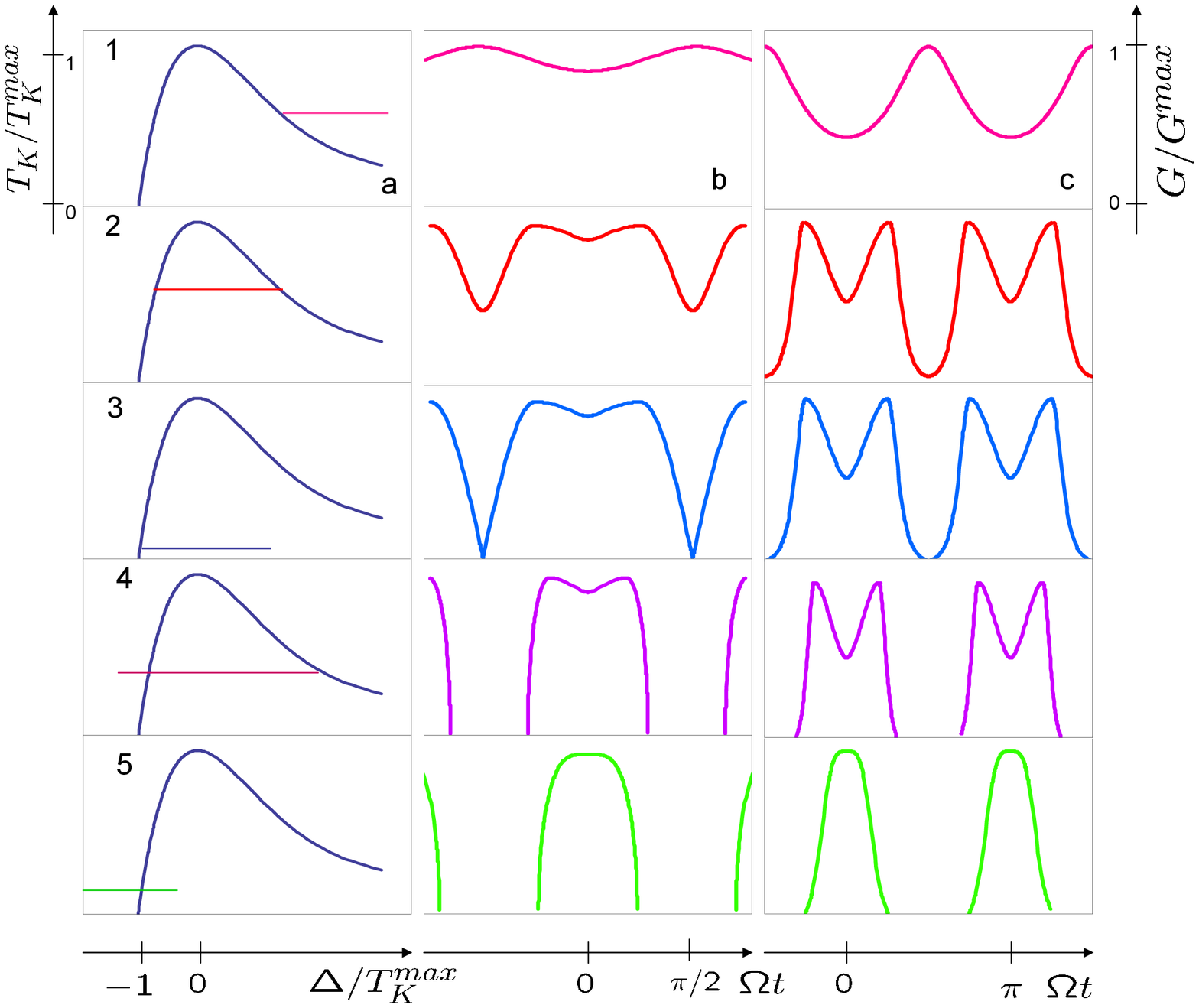}
\end{center}
\vspace*{-2cm} \caption{(Color online) Left column: $T_K$ as a
function of $\Delta_{ST}$. Middle column: time dependent $T_K(t)$
corresponding to the evolution of $T_K(\Delta_{ST})$ in the left
column. The intervals over which the evolution is followed are
shown by straight lines in the left column. Right column: Evolution
of ZBA in the conductance according to Eq. (\ref{conduct}). See the
text for further discussion. }\label{f.2}
\end{figure}

It is known from Kondo theory for quantum dots with $SO(4)$
symmetry,\cite{KA01,Eto,Pust1} that the Kondo temperature is a
very sharp function of $\Delta_{ST}$ with a maximum $T_{K0}$ at
$\Delta_{ST}=0$ (see  Fig. \ref{f.2}, left column). It has an
intermediate asymptotic behaviour for positive $\Delta_{ST}$ where
the ground state of DQD is a triplet
\begin{equation}
\label{asymp}
    \frac{T_K(t)}{T_{K0}}=\left[\frac{T_{K0}}{\Delta_{ST}(t)}
    \right]^\eta,
    \end{equation}
valid in an intermediate asymptotic regime for positive $\Delta_{ST}$
at $T_{K0}/\Delta_{ST}\lesssim 1$. Here $\eta<1$ is a universal
constant. This sharp dependence is a key to the charge-spin
transformation mechanism, which is especially effective in the
vicinity of the triplet/singlet critical point $\Delta_{ST}=0$.

Next, we estimate the influence of the trembling signal on the
tunnel conductance $G(T,t)$ at given $T>T_K$ in a situation where
the temporal variations of $\Delta_{ST}(t)= E_S-E_T - \delta_{\rm
coh} (t)$  changes $T_K (t)$ close to the above mentioned
degeneracy point $\Delta_{ST}=0$. In this high-temperature weak
coupling regime the zero bias anomaly (ZBA) in tunnel conductance
obeys the law
\begin{equation}\label{conduct}
\frac{G}{G_0}\sim \ln^{-2}(T/T_K).
\end{equation}
Substituting (\ref{asymp}) in (\ref{conduct}), one gets
\begin{equation}\label{conducttt}
\frac{G(t)}{G_0} \sim\left(\ln(T/T_{K0}-\eta\ln
(T_{K0}/\Delta_{ST}(t) \right)^{-2}
\end{equation}

The results of the numerical analysis of the Kondo response $T_K(t)$
to the periodical input given by Eq. (\ref{pergate}) as well as
the adiabatically varying conductance (\ref{conduct}) are
presented in Fig. \ref{f.2}. It is seen from this analysis that
the input sinusoidal potential $\tilde v_g\cos \Omega t$ transforms
into periodic oscillations of the Kondo-related ZBA in  $G(t)$. The
sinusoid is reproduced with a slight distortion
 when the gap $\Delta_{ST}(t)$ remains positive under
temporal perturbation (row 1 in Fig. \ref{f.2}). Additional minima
impart the "Kremlin wall" shape to the periodic curve $G(t)$
when the sign of $\Delta_{ST}(t)$ changes under the periodic
perturbation (rows 2,3). The same regime with larger amplitude
$\tilde v_g$ may result in complete suppression of Kondo tunneling
due to the periodical triplet-singlet crossover (row 4). Finally,
if the system remains completely in the singlet sector
$\Delta_{ST}(t)<0$ near the crossover point, the charge
perturbation results in a pulsed Kondo output signal (row 5).
 \smallskip\\

 It is worth noticing
that this mechanism of adiabatic transformation of a charge signal
into a Kondo response is close to that proposed for Kondo shuttling
\cite{shut} where the source of time-dependence are the
nanoelectromechanical oscillations of a quantum dot with even
occupation between two leads.

The reference Kondo temperature $T_{K0}$ is given by the equation
\begin{equation}\label{3.1}
T_{K0}= \bar D e^{-\frac{1}{\left(j^T_0 + j^S\right)}}
\end{equation}
where $j^\Lambda = \rho_F J^\Lambda$, $\rho_F$ is the density of
electron states in the leads at the Fermi level, and $\bar D$ is
the characteristic scale of these states in SW regime.\cite{KA01}
This temperature also oscillates adiabatically in time due to the
correction (\ref{josc}) to the second term in the exponent. One
may estimate adiabatic oscillations  $\delta T_{K0}^{\rm ad}$ in
the lowest order in $j^S_{\rm ad}/(J^T_0 + J^S_0)\ll 1$. One
derives from (\ref{3.1})
\begin{equation}\label{3.2}
\delta T_{K0}^{\rm ad}(t) \approx T_{K0} \frac{j^S_{\rm
ad}(t)}{(j^T_0 + j^S_0)^2} \sim
T_{K0}\frac{\widetilde\phi_1(t)}{j^T_0}~.
\end{equation}
These temporal oscillations only weakly distort the curves shown
in Fig. \ref{f.2} because the Kondo temperature in these curves
changes by the order of its magnitude due to oscillations changing
the sign of $\Delta_{ST}(t)$. One may roughly estimate this effect
by averaging (\ref{3.2}) over an oscillation period. Like in the
Kondo shuttling,\cite{shut} this averaging results in an effective
enhancement of $T_{K0}$ which resembles the Debye-Waller
enhancement of neutron scattering intensity because of a nonzero
average quadratic displacement induced by phonon vibrations.

\subsection{Stochastic input}

 We begin with the discussion of the influence of an
incoherent input on the spin state of  quantum dot {\it isolated}
from a metallic reservoir. As is known from the general theory of
dynamical symmetries, \cite{Nova} only those states from the total
manifold are involved in its formation whose energies are
comparable with the energy scale of the interaction which breaks
the symmetry of the Hamiltonian. In our case this scale is
determined by the Kondo temperature $T_K \sim \Delta_{ST}\ll
\Delta_{ET}$.

\subsubsection{Fluctuations of the global constraint }

The mechanism of conversion of the stochastic component $\delta
v_g(t)$ of the input signal into a stochastic spin response is
quite unusual. Instead of dephasing due to time-dependent spin
flip processes, \cite{KNG} stochastization of the energy spectrum
of DQD results in the loss of a Curie-type spin response at some
characteristic energy $\zeta$. This effect is related to the time
dependence of the factors $\delta_{{\rm stoch},\Lambda}(t)$
(\ref{delta}) in the Hamiltonian (\ref{hdott22}). Indeed,
inserting (\ref{stadia}) into (\ref{hdott22}), one may write the
stochastic part of $H_{\rm dot}$ in the form
\begin{eqnarray}\label{cost}
H_{\rm dot}^{\rm stoch}= [\delta_{{\rm stoch},S}(t){\bf P}^2 -
\delta_{{\rm stoch},E}(t){\bf M}^2]/2
\end{eqnarray}
Unlike the adiabatic part of time dependent energies $E_\Lambda(t)$,
this term describes spin fluctuations related to the dynamical
symmetry of DQD. In the reduced singlet/triplet subspace the exact
confinement preserved by the last term in (\ref{hdott22}) obeying
$SO(5)$ dynamical symmetry transforms into fluctuating confinement
in the effective Hamiltonian,
\begin{eqnarray}\label{constrab}
H_{dot}(t)&=& \frac{1}{2}\left(E_T{\bf S}^2 + E_S{\bf P}^2 \right)\\
&-&\mu({\bf S}^2 + {\bf P}^2-3) +\delta_{{\rm stoch},S}(t){\bf
    P}^2 \nonumber
\end{eqnarray}
where the $SO(4)$ symmetry is preserved only approximately. Thus
the stochastic component of the dot Hamiltonian given by the
correlation functions (\ref{delta}) appears explicitly in the
constraint.

It follows from (\ref{constrab}) that this component survives even
in the asymptotic regime, $T \ll \Delta_{ST}$, where the T/S
excitations are quenched in the Kondo scattering, but the singlet
component of the spin multiplet still influences the constraint
via its stochastic constituent. The effective dot Hamiltonian in
this limit has the form
\begin{eqnarray}\label{constrac}
H_{dot}(t)= \frac{1}{2}E_T{\bf S}^2  -\mu({\bf S}^2 -2)
-\delta_{{\rm stoch},S}(t){\bf S}^2~.
\end{eqnarray}
so that the fluctuations of the \textit{charge} S/E gap may be
transformed into the fluctuations of the \textit{spin} constraint.
This unusual situation is considered below in greater detail.

To investigate the influence of stochastic corrections of $\mu$ on
the spin properties of isolated dot, we rewrite the Hamiltonian
(\ref{constrac}) in a fermionized form
\begin{equation}\label{hferm}
H_{dot}(t)= \sum_{\nu=0,\pm 1} [E_T/2 - \mu(t)]f^{\dag}_\nu
f^{}_\nu .
\end{equation}
Here $f_{\nu}$ are spin fermions representing the S=1
triplet,\cite{Nova,KKAR} and the time-dependent chemical potential
for spin fermions is defined as $\mu(t)= \mu_0 - \delta_{{\rm
stoch},S}(t)$. The stochastic component of $\mu$ may be treated as
a random potential in the time domain, which describes the fluctuations
of global fermionic constraint.\cite{KKAR} The problem of
propagation of spin fermions in a random time-dependent potential
may be considered by means of the "cross technique"\cite{Edw}
developed for the study of electron propagation in a field of
impurities randomly distributed in real space.

In Ref. \onlinecite{KKAR} the short-time (white-noise)
fluctuations of a global fermionic constraint have been considered.
The noise correlation in this limit is delta-like
\begin{equation}\label{whno}
D(t-t')=\hbar^2\langle\mu(t)\mu(t')\rangle =r_0\delta(t-t').
\end{equation}
     Such a description presumes that the chemical
potential suddenly  "shaken" at any moment does not keep memory
about its previous value (correlation time equals zero). The
decoherence time calculated in Born approximation is given by
$$
\hbar/\tau_d\sim r_0
$$

Here we propose another realization of the stochastic potential,
which corresponds to the situation when the chemical potential
varies slowly in time ($\sim \exp(-\gamma t)$). A very long
relaxation time $\tau_r\sim 1/\gamma$ with small $\gamma$ is
assumed, so that the noise correlation is given by
\begin{equation}\label{fluct}
D(\omega)=\lim_{\gamma\to 0}
\frac{2\zeta^2\gamma}{\omega^2+\gamma^2}=2\pi
\zeta^2\delta(\omega)
\end{equation}
In this limit the averaged spin propagator describes the ensemble
of states with  chemical potential $\mu =const$ in a given state,
but this constant is random in each realization.\cite{footkayan}

The problem of decoherence of the spin state in a stochastically
perturbed DQD in this limit can be mapped on the so-called Keldysh
model \cite{keld65,efros70,sad} originally formulated for systems
which are $\delta$-correlated in the momentum space impurity scattering
potential. The problem can be solved exactly and the decoherence
time is defined by the variance $\zeta^2$ of the Gaussian
correlation (see below). We look for a solution of the
time-dependent model where time is the only current coordinate in
the system.

\begin{figure}[h]
  \includegraphics[width=1.5cm,angle=0]{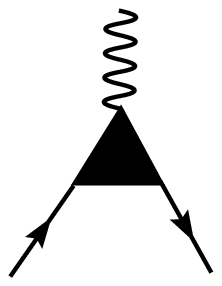}\hspace*{10mm}
  \includegraphics[width=1cm,angle=0]{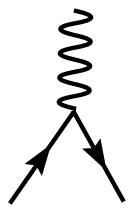}\hspace*{10mm}
  \includegraphics[width=1.5cm,angle=0]{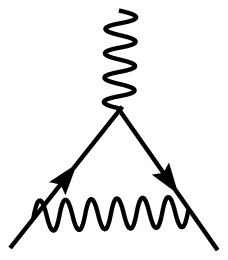}
  \caption{Diagrams for the vertex $\Gamma$.}\label{f.vert1}
\end{figure}

The spin-fermion propagator at $T=0$ is defined as
\begin{equation}\label{get}
G^R_{T\nu}(t-t')=\langle
f^{}_\nu(t)f^\dag_\nu(t')\rangle_R=-i\langle[f^{}_\nu(t)f^\dag_\nu(t')]_+\rangle.
\end{equation}
We sum the perturbation series for the Fourier transform of this
Green's function
\begin{equation}\label{seriess}
G^R(\varepsilon) = g(\varepsilon)\left[1+\sum_{n=1}^\infty A_n
\zeta^{2n}g^{2n}(\varepsilon)
 \right]
\end{equation}
Here $g(\varepsilon)=(\varepsilon+i\delta)^{-1}$ is the free
spin-fermion propagator with $E_T/2-\mu_0=0$ taken as the reference
energy. The index $\nu$ is omitted, since the fluctuations
of $\mu$ are related to the global $U(1)$ symmetry.
  The noise correlation function (\ref{fluct}) is normalized  in such a way
that corresponding vertices are dimensionless.
\begin{figure}[h]
  \includegraphics[width=2cm,angle=0]{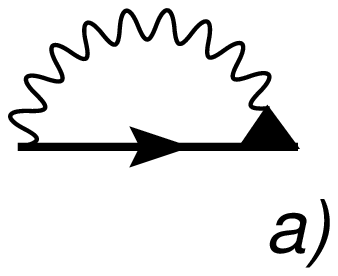}\hspace*{10mm}
  \includegraphics[width=2cm,angle=0]{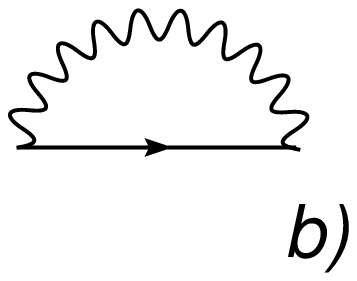}\\\vspace*{5mm}
  \includegraphics[width=2cm,angle=0]{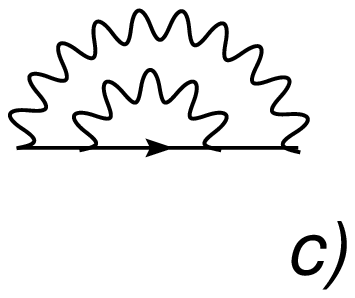}\hspace*{10mm}
  \includegraphics[width=2cm,angle=0]{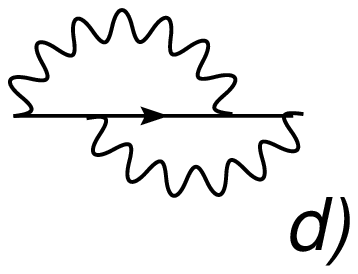}\\ \vspace*{8mm}
  \caption{(a) Feynman diagrams for the self energy with vertex corrections;
  (b) Bare self energy; (c) First line correction;
  (d) First vertex correction.}\label{f.self1}
\end{figure}
The Feynman diagrams for the self-energy $\Sigma(\epsilon)$ are
shown in Fig. \ref{f.self1}. The key features of the Keldysh model
stem from the fact that all self energy diagrams in a given order
are equivalent due to the delta-function character of the
correlation function $D(\omega)$ (\ref{fluct}). As a result, the
contribution of all diagrams in a given order $n$ is completely
determined by the combinatorial coefficient $A_n =(2n-1)!!$ giving
the total number of diagrams corresponding to all possible
pairwise connections of $n$ vertices by wavy lines. Then the exact
analytical equation for the self energy may be derived,
\cite{keld65,efros70,sad}
\begin{equation}
\Sigma(\epsilon)=\int\frac{d\omega}{2\pi}
\Gamma(\epsilon,\epsilon-\omega;\omega)G(\epsilon-\omega)D(\omega)
\label{sigma}
\end{equation}
Here $\Gamma$ is the full vertex (triangle), $G$ is the full Green
function (thick line)  and $D$ is the noise correlation function
(wavy line) in Fig. \ref{f.self1}a. Evaluation of the integral
(\ref{sigma}) with the $\delta$-functional $D(\omega)$
(\ref{fluct}) gives
\begin{equation}
\Sigma(\epsilon)=\zeta^2\Gamma(\epsilon,\epsilon;0)G(\epsilon)
\label{sgm}
\end{equation}
In order to  find $G(\epsilon)$ we use the Ward identity illustrated
by Fig. \ref{f.vert1}) which connects the triangular vertex and the
Green's function (GF)
\begin{equation}
\Gamma(\epsilon,\epsilon;0)=\frac{dG^{-1}(\epsilon)}{d\epsilon}
\label{vert}
\end{equation}
Then the Dyson equation for the spin-fermion propagator is transformed
with the help of (\ref{vert}) into an ordinary differential
equation:
\begin{equation}
\zeta^2\frac{d G}{d \epsilon}+ \epsilon G -1=0 \label{DE}
\end{equation}
 This equation is supplemented by the
boundary condition \cite{efros70,sad}
\begin{equation}
G(\epsilon\to \infty)=\frac{1}{\epsilon} \label{ass}
\end{equation}
The solution of (\ref{DE}) satisfying the boundary condition
(\ref{ass}) is given by
\begin{equation}
G^R(\epsilon)=\frac{1}{\zeta\sqrt{2\pi}}\int_{-\infty}^{\infty}e^{-z^2/2\zeta^2}
\frac{dz}{\epsilon-z+i\delta} \label{GR}
\end{equation}
Remarkably, the spin-fermion GF in this model has no poles,
singularities or branch cuts. The solution (\ref{GR}) represents the
set of spin states under a stochastically fluctuating chemical
potential averaged with a Gaussian exponent characterized by the
variance $\zeta^2$.

\begin{figure}[h]
  \includegraphics[width=2cm,angle=0]{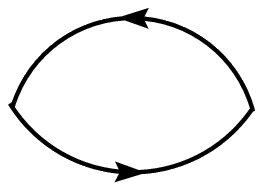}\hspace*{5mm}
  \includegraphics[width=2cm,angle=0]{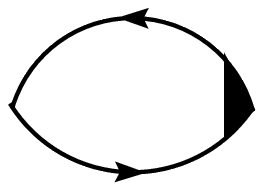}
  \caption{Diagrams for bare and dressed spin susceptibility.}\label{f.loop1}
\end{figure}

Let us investigate the spin response of this "stochasticized" DQD.
To calculate the spin susceptibility, it is convenient to make an
analytic continuation of $G^R$ on the imaginary semi-axes of
complex energies, i.e. to go over to the thermodynamical Matsubara
Green's functions:
\begin{equation}
{\cal G}
(i\epsilon_n)=\frac{1}{\zeta\sqrt{2\pi}}\int_{-
\infty}^{\infty}e^{-z^2/2\zeta^2}\frac{dz}{i\epsilon_n-z}
\label{GGR}
\end{equation}
The spin (triplet) susceptibility at finite temperatures defined
by the diagrams of Fig. \ref{f.loop1} may be calculated by means
of these functions.
The spin susceptibility with vertex corrections is determined as
\begin{eqnarray}\label{chi}
&&\chi(i\omega_m)=\\
&& T\sum_n{\cal G}(i\omega_m+i\epsilon_n){\cal
G}(i\epsilon_n)\Gamma(i\epsilon_n,i\epsilon_n+i\omega_m;i\omega_m)\nonumber
\end{eqnarray}

The Ward identity (\ref{vert}) provides us with the exact equation
for the vertex function
\begin{equation}
\Gamma^R(\epsilon,\epsilon;0)=\frac{\epsilon
G^R-1}{\zeta^2(G^R)^2}, \label{GamR}
\end{equation}
giving access to the exact evaluation of the static susceptibility
$\chi(0)$ (right diagram on Fig. \ref{f.loop1}). Combining
(\ref{chi}) with (\ref{GGR}) and (\ref{GamR}), we find
\begin{equation}
\chi(0)=\frac{1}{\sqrt{8\pi}\zeta}\int_{-\infty}^{\infty}dy
e^{-y^2/2} y\tanh\left(\frac{y\zeta}{2T}\right) \label{ssuca}
\end{equation}

 The asymptotic behavior of the static susceptibility
$\chi(0)$ is
\begin{eqnarray}
\chi(0) =\left\{
\begin{array}{c}
C_{\rm C}/T,\;\;\; T\gg \zeta\\
\\\label{Gm}
C_{\rm K}/\zeta,\;\;\; \zeta\gg T\\
\end{array}\right.
\end{eqnarray}
where $C_{\rm C}$ and $C_{\rm K}$ are constants. At high $T$ the
behavior of the dot is Curie-like with Curie constant $C_{\rm C}$
modified by averaging. At low $T$ the noise dispersion $\zeta$
plays the role of an effective temperature in the Keldysh model
with the corresponding constant $C_{\rm K}$ in the numerator.

There is a great simplification in the calculations of the dynamic
susceptibility at temperatures $T\gg \zeta$. We notice that
$\Gamma\to 1$ in this limit since transferred energy exceeds the
dispersion of the noise spectrum. Under this condition one can
neglect the vertex corrections, and the spin susceptibility is
given by
$$
\chi(i\omega_m)= \frac{1}{2\pi
\zeta^2}\int_{-\infty}^{\infty}\frac{dz_1dz_2
e^{-\frac{z_1^2+z_2^2}{2\zeta^2}}}{\cosh(\frac{z_1}{2T})\cosh(\frac{z_2}{2T})}
\frac{\sinh\left(\displaystyle\frac{z_2-z_1}{2T}\right)}{i\omega_m+z_2-z_1}
$$
Performing the analytic continuation, one gets the following
equation for the imaginary part of the spin response function at
real frequencies:
\begin{eqnarray}\label{imchi}
&& Im \chi^R(\omega)=
\\
&&-\frac{e^{-\frac{\omega^2}{4\zeta^2}}}{\zeta^2}\tanh\left(\frac{\omega}{2T}\right)
\int_{-\infty}^{\infty}d q\displaystyle\frac{
e^{-\frac{q^2}{4\zeta^2}}}{\displaystyle
\frac{\cosh(\frac{q}{2T})}{\cosh(\frac{\omega}{2T})}+1}\nonumber .
\end{eqnarray}
Thus $Im \chi^R\sim \omega$ at small $\omega\ll T$.

The real part of the static susceptibility is given by
\begin{eqnarray}\label{rechi}
&&Re \chi^R(\omega)= \\
&&\frac{1}{2\pi \zeta^2}
\int_{-\infty}^{\infty}du\frac{u\sinh\left(\frac{u}{2T}\right)}{u^2-
\omega^2}e^{-\frac{u^2}{4\zeta^2}}\times\nonumber\\
&&\times \int_{-\infty}^{\infty}d q
\frac{e^{-\frac{q^2}{4\zeta^2}}}{\cosh(\frac{u}{2T})+\cosh(\frac{q}{2T})}\nonumber
\end{eqnarray}
where the principal part of the integrals is taken.
\begin{figure}[h]
\includegraphics[width=4cm,angle=0]{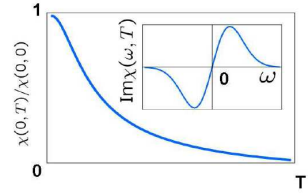}
\caption{(Color online) Static susceptibility $\chi(0,T)$. The inset
shows a frequency dependence of ${\rm Im}\chi$ with a maximum at
$\omega \sim \zeta$.}\label{f.44}
\end{figure}

It follows from (\ref{ssuca}), (\ref{imchi})  and (\ref{rechi})
and from results of numerical calculations presented in Fig.
\ref{f.44} that the low-frequency response of a DQD in the Keldysh
regime has nothing to do with the behavior of a free spin. This
means that in spite of the fact that at high $T$ a DQD behaves
like a quantum object with spin 1, it looses at $\{\omega,T\}\ll
\zeta$ the generic characteristics of a localized spin due
stochastization, hence it cannot serve as a source of Kondo screening
at low energies.

\begin{figure}[h]
  \includegraphics[width=2cm,angle=0]{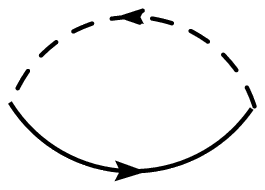}\hspace*{5mm}
  \includegraphics[width=2cm,angle=0]{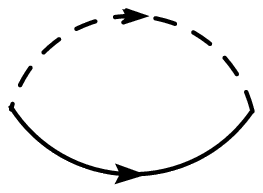}
  \caption{Diagrams for spin-electron loops responsible for the Kondo effect.
  Here solid and dashed lines stand for spin-fermion propagators
  ${\cal G}(i\epsilon_n)$
  and electron propagators $g({\bf p},i\omega_n)$, respectively.}
\label{f.loop2}
\end{figure}

A direct calculation of spin-electron loops responsible for Kondo
screening (Fig. \ref{f.loop2}) confirms this conclusion. These
second order perturbation theory corrections to the spin-electron
vertex are given by
$$
{\cal T}^{(2)} (i\epsilon_n)/J \sim J T\sum_m \int \frac{d {\bf
p}}{(2\pi)^3}{\cal G}({\bf p},i\omega_m){\cal
G}(i\epsilon_n+i\omega_m)
$$
Substituting Green's functions into this integral, we have
$$
{\cal T}^{(2)} (i\epsilon_n)/J \sim (\rho J)
\int_{-\infty}^{\infty}dze^{-\frac{z^2}{2\zeta^2}}\int_T^D
d\xi\frac{\displaystyle\tanh\left(\frac{\xi}{2T}\right)}{i\epsilon_n-\xi+z}
$$
This vertex correction reflects the averaging of the standard
Kondo vertex with the Gaussian distribution of chemical
potentials. Evaluating the integral, one obtains a combination of
logarithmic, hypergeometric and imaginary error functions
\vspace*{-1mm}
\begin{equation}\label{3.3}
{\cal T}^{(2)}(\epsilon\to 0)/J=\rho_0 J\ln(\sqrt{2C}D/\zeta)+
\end{equation}
\vspace*{-7mm}
\begin{eqnarray}
+\frac{1}{2}\rho_0
J\left[_1F_1\left(\frac{3}{2},2,\frac{T^2}{2\zeta^2}\right)\left(\frac{T}{\zeta}\right)^2
-\pi {\tt Erfi}\left(\frac{T}{\sqrt{2}\zeta}\right)\right]
\nonumber
\end{eqnarray}
where $\gamma=\ln C$ is the Euler constant. In
two limiting cases of low  and high temperatures relative to the
dispersion $\zeta$ of noise spectrum,  it leads to the following
compact expressions
\begin{eqnarray}
{\cal T}^{(2)} (\epsilon\to 0)/J=\left\{
\begin{array}{c}
\rho J\ln(D/T),\;\;\; T\gg \zeta\\
\\\label{Gma}
\rho J\ln(D/\zeta),\;\;\; \zeta\gg T\\
\end{array}\right.
\end{eqnarray}
We conclude from here that the variance $\zeta^2$ predetermines
the energy/temperature cut-off (similarly to the Kondo-spin glass
problem \cite{kisop00}). This result correlates with the above
observation that at low $T\ll \zeta$ the magnetic excitations in
stochasticized DQD loose the properties of spin flip processes
which are essential for Kondo screening. The characteristic
decoherence time is given by:
$$
\hbar/\tau_d \sim \zeta
$$
If $\zeta\ll T_K$, the noise effect is seen in the behavior of the
magnetic susceptibility as a logarithmic correction,
$$
\chi(T) = \chi_C\left(1-\ln^{-1}(\max(T,\zeta)/T_K)+...\right),
$$
but to study the influence of $\delta v_g$ (\ref{1.3o}) on the
Kondo processes at finite temperatures one should also consider
the dephasing effects.

We comment also the S-T response function
$$
\chi^{\alpha\beta}_P(t-t')=-i\langle [P_\alpha(t),
P_\beta(t')]\rangle\to\delta_{\alpha\beta}\chi_P(i\omega_n)
$$
which corresponds to the bare loop represented in Fig.
\ref{f.loop1}a, where one of the two lines corresponds to the singlet
fermionic GF while the other one represents the triplet fermionic
GF. Since the singlet line is not affected by the noise, only one
of the GFs in the loop suffers from Gaussian averaging.

The straightforward calculations lead to the following answer for
the imaginary part
\begin{eqnarray}
&& Im \chi^R_P(\omega)\sim
\\
&&-\frac{1}{\zeta}
\frac{\sinh\left(\frac{\omega}{2T}\right)}{\sinh\left(\frac{\Delta_{ST}}{2T}\right)}
\left[\frac{\exp(-\frac{(\Delta_{ST}+\omega)^2}{2\zeta^2})}{\cosh(\frac{\Delta_{ST}+\omega}{2T})}
+\frac{\exp(-\frac{(\Delta_{ST}-\omega)^2}{2\zeta^2})}{\cosh(\frac{\Delta_{ST}-\omega}{2T})}\right]
\nonumber
\end{eqnarray}
and the real part of the susceptibility
\begin{eqnarray}
&& Re \chi^R_P(\omega)\sim
\\
&&\frac{1}{\zeta}
\int_{-\infty}^{\infty}dz\frac{(z-\Delta_{ST})\sinh\left((z-\Delta_{ST})/2T\right)}{(z-\Delta_{ST})^2
-\omega^2}\times\nonumber\\
&&\times \frac{e^{-\frac{\displaystyle z^2}{
2\zeta^2}}}{\cosh(z/2T)\cosh(\Delta_{ST}/2T)} \nonumber
\end{eqnarray}

The static susceptibility which mimics the Curie law at very large
temperatures $T\gg(\Delta_{ST}, \zeta)$, is suppressed
exponentially $\sim\exp(-\Delta_{ST}/T)$  at low temperatures
$T\ll (\Delta_{ST}, \zeta)$ and has an intermediate asymptotic behaviour
$\chi_P(0)\sim 1/\zeta\ln(\zeta/T)$ if $\Delta_{ST}\ll T\ll \zeta$
while $\chi_P(0)\sim 1/\Delta_{ST}$ when $\zeta\ll T\ll
\Delta_{ST}$. The real part of the dynamic susceptibility taken at
the resonance frequency $\omega=\pm\Delta_{ST}$ grows as
$1/\zeta\ln(\zeta/T)$ when the temperature is lowered. The
exponential suppression does not occur since $\Delta_{ST}$ is
compensated by the external frequency. We therefore conclude
that the noise may strongly affect the non-equilibrium
electric-field-induced Kondo transport in the regime when the
singlet is a ground state of DQD while the access to the triplet
state is facilitated by the applied gate voltage.

\subsubsection{Fluctuations of the scattering phase}

\begin{figure}[h]
  \includegraphics[width=5cm,angle=0]{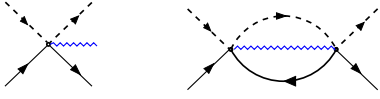}\hspace*{5mm}
    \caption{Left: bare vertex in the fluctuation part of the exchange Hamiltonian
    (\ref{hscatt}).
    Right: first non-adiabatic correction to the exchange vertex $J^{T}_0$ in
    (\ref{SW2}).}\label{f.scatter}
\end{figure}

In accordance with the general approach to decoherence and dephasing
effects,\cite{Zeh} the latter phenomena arise due to scattering
processes at finite energy and/or temperature. These processes are
described by the effective cotunneling Hamiltonian (\ref{SW2}).
The main contribution to dephasing is given by the term
\begin{equation}\label{hscatt}
H^{(\rm stoch)}_{\rm cotun}= J_{\rm stoch}^{S}(t) {\bf P}\cdot
{\bf s}
\end{equation}
Here $J_{\rm stoch}^{S}(t)$ is the stochastic component of the
indirect exchange integral calculated in Appendix A (Eq.
\ref{Apex}). To reveal the dephasing mechanism, one should notice
that the parameter $\overline {J^S_{\rm st}(t)}$
\begin{equation}\label{stex}
\overline {J^{S}_{\rm st}(t)} = \frac{W^{2}}{\epsilon_2-M_S}
\frac{2V}{\Delta_{ES}}
\left(\frac{\overline{v_{g}(t)\phi_1(t)}}{\epsilon_2} -
\overline{\phi_1(t)^{2}}\right)
\end{equation}
is in fact the modification of the effective SW exchange due to
temporal fluctuations of the intradot exchange $2\alpha V$, which is
nothing but the gap $\Delta_{ST}$ (see Eq. \ref{levels}). Unlike
the similar term $\delta_{\rm stoch}(t)$ in Eqs. (\ref{stadia}),
(\ref{constrab}), it does not influence the modulus of the vector $\bf
P$, but the components of this vector, and thereby it affects the
components of the spin vector ${\bf S}$ via scattering processes
illustrated by Fig. \ref{f.scatter}. The time-dependent exchange
vertex (\ref{stex}) is taken in the form $ J_{so}\varphi_s(t)$, so
that the fluctuating part in parenthesis is represented by a
single mode $\varphi(t)$.

\noindent The wavy lines in the diagram stand for the correlation
functions $S(t-t')=\langle\varphi(t)\varphi(t')\rangle$, the solid
line in the vertex correction corresponds to the bare retarded
propagator $g_s(t-t')=\langle f^{}_s(t)f^\dag_s(t')\rangle_R$ for
pseudofermion excitations representing the singlet mode in the
$SO(5)$ multiplet. \cite{Nova,KA01} As was noted in I, the
eventual source of dephasing is the gauge fluctuation field
induced by non-adiabatic excitations in the non-diagonal operator
${\bf P}$ with components $P_\nu = f^\dag_s f^{}_{\nu}$.

The problem of dephasing due to slow fluctuations $S(\omega\to 0)$
was analyzed in I, so we do not repeat here the corresponding
calculations. The net result of this analysis is that the
dephasing processes  are relevant at high enough temperatures, and
at $T\ll \Delta_{ST}$ dephasing is effectively quenched.

\section{Triple quantum dot in a triangular geometry. Odd electron
occupation}

In this section we sort out the coherent and stochastic components
of the weak probe time-dependent potential $v_g(t)$ (\ref{1.3o})
applied to the triangular TQD shown in Fig. \ref{f.1}b. To make the
mathematical treatment more transparent, we consider the
three-terminal geometry, so that the system possesses a perfect
triangular symmetry $C_{3V}$. It is convenient to enumerate both
the dots and the leads by the numbers 1,2,3. The bias is supposed
to be created between the leads 2,3, and the role of a passive
terminal 1 is to serve as a reservoir for Kondo screening of the
electron spin in the dot. The gate voltage $v_g(t)$ is applied to
one of the electrodes forming the dot 1.

Then the band Hamiltonian has the form
\begin{equation}\label{hbandt}
H_{band}=\sum_{j=1,2,3}\varepsilon_{jk}
c^\dag_{jk\sigma}c^{}_{jk\sigma}~,
\end{equation}
Correspondingly, the tunnel Hamiltonian is written as
\begin{equation}\label{htunt}
H_{tun}=W\sum_{jk\sigma}( c^\dagger_{jk\sigma}d^{}_{j\sigma}+
H.c.),
\end{equation}
We study the excitation spectrum of the TQD in the charge sector
${\cal N}=1$. The spin degrees of freedom obey the $SU(2)$
symmetry, and all dynamical symmetry effects are related to {\it
orbital} degrees of freedom. We will show here how the influence
of charge input on the orbital degrees of freedom may be converted
into Kondo response.

The spectrum of the TQD  was discussed in Refs.
\onlinecite{KKA,kork}. This dot is described by the Hamiltonian
\begin{eqnarray}\label{tqd}
&&H_{dot}^{(0)}=\epsilon\sum_{j=1}^3\sum_{\sigma}d^\dagger_{j
\sigma}d_{j\sigma }+Q\sum_{j}n_{j\uparrow}n_{j\downarrow}\label{H-dot} \\
&& +Q'\sum_{\langle jl\rangle}\sum_\sigma n_{j\sigma}n_{l\sigma'}
+V\sum_{\langle jl\rangle}\sum_{\sigma}(d^\dagger_{j
\sigma}d_{l\sigma }+H.c.). \nonumber
\end{eqnarray}
Here $n_{j \sigma}=d^\dagger_{j \sigma}d_{j\sigma },$ $\langle jl
\rangle= \langle12\rangle,\langle23\rangle,\langle31\rangle$, $Q$
and $Q'$ are intradot and interdot
 charging energies ($Q\gg Q'$), $V$ is the interdot tunneling
 amplitude. In the case ${\cal N}=1$ charging terms are irrelevant and
 (\ref{tqd}) is easily diagonalized
\begin{equation}
H_{dot}^{(0)}=\sum_{\Gamma,\sigma} \varepsilon_\Gamma
d^\dag_{\Gamma\sigma} d^{}_{\Gamma\sigma}
\end{equation}
Here the index $\Gamma=A,E_\pm$ stands for irreducible
representations of the symmetry group of equilateral triangle. The
basis of this representation is given by the eigenfunctions
\begin{eqnarray}\label{A1}
    &&d^\dag_{A,\sigma}=
(d^\dag_{1\sigma}+d^\dag_{2\sigma}+d^\dag_{3\sigma})/\sqrt{3}~,\\
&&d^\dag_{E_\pm,\sigma}= (d^\dag_{1\sigma}+e^{\pm
2i\varphi}d^\dag_{2\sigma}
    +e^{\pm i\varphi}d^\dag_{3\sigma})/\sqrt{3}~; \nonumber
\end{eqnarray}
$(\varphi=2\pi/3)$ The ground state is the orbital singlet $A$,
and the orbital degrees are quenched at low temperatures $T\ll V$.

\begin{figure}[h]
\includegraphics[width=4.2cm,angle=0]{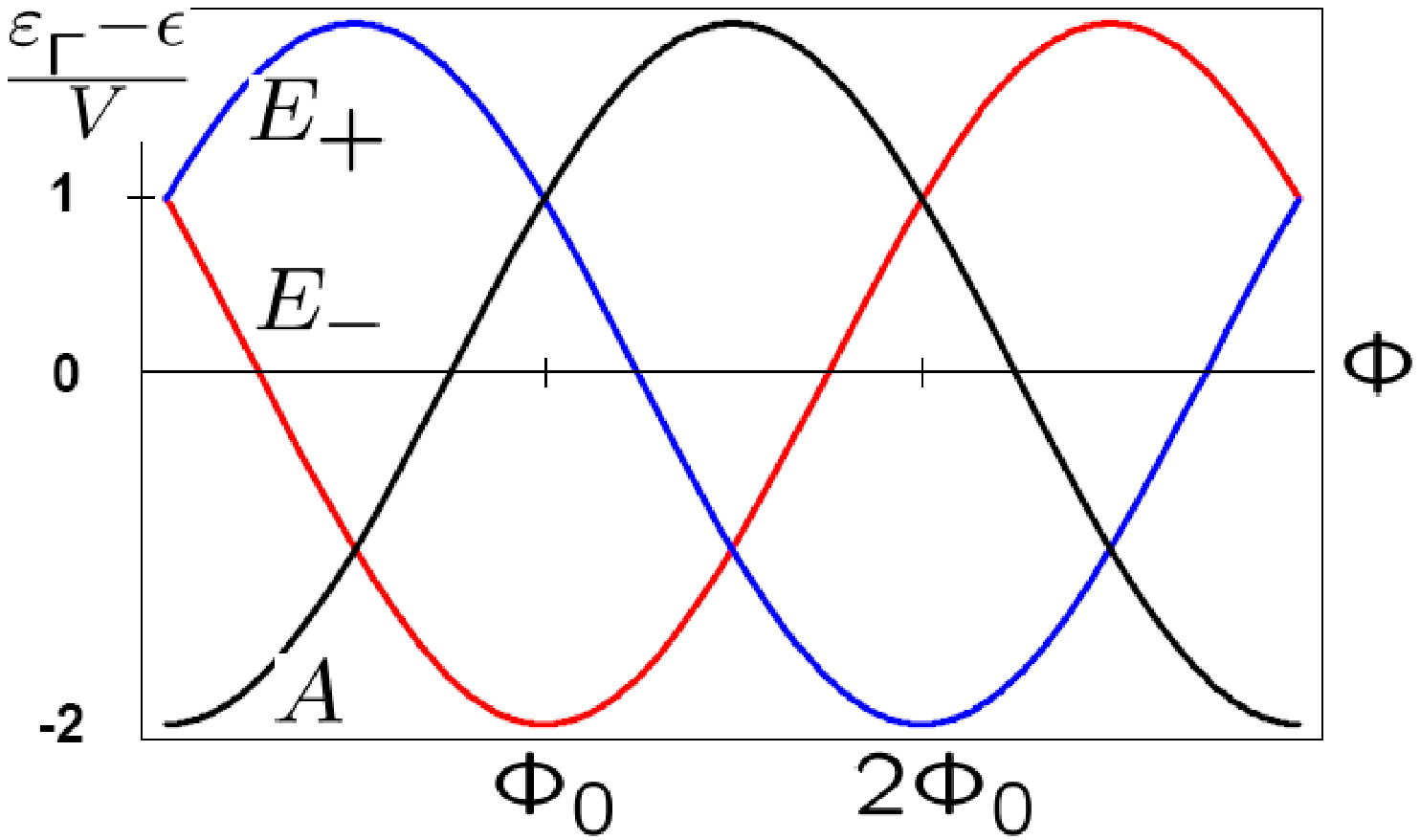}\hspace*{12mm}
\includegraphics[height=2.55cm,width=3.cm,angle=0]{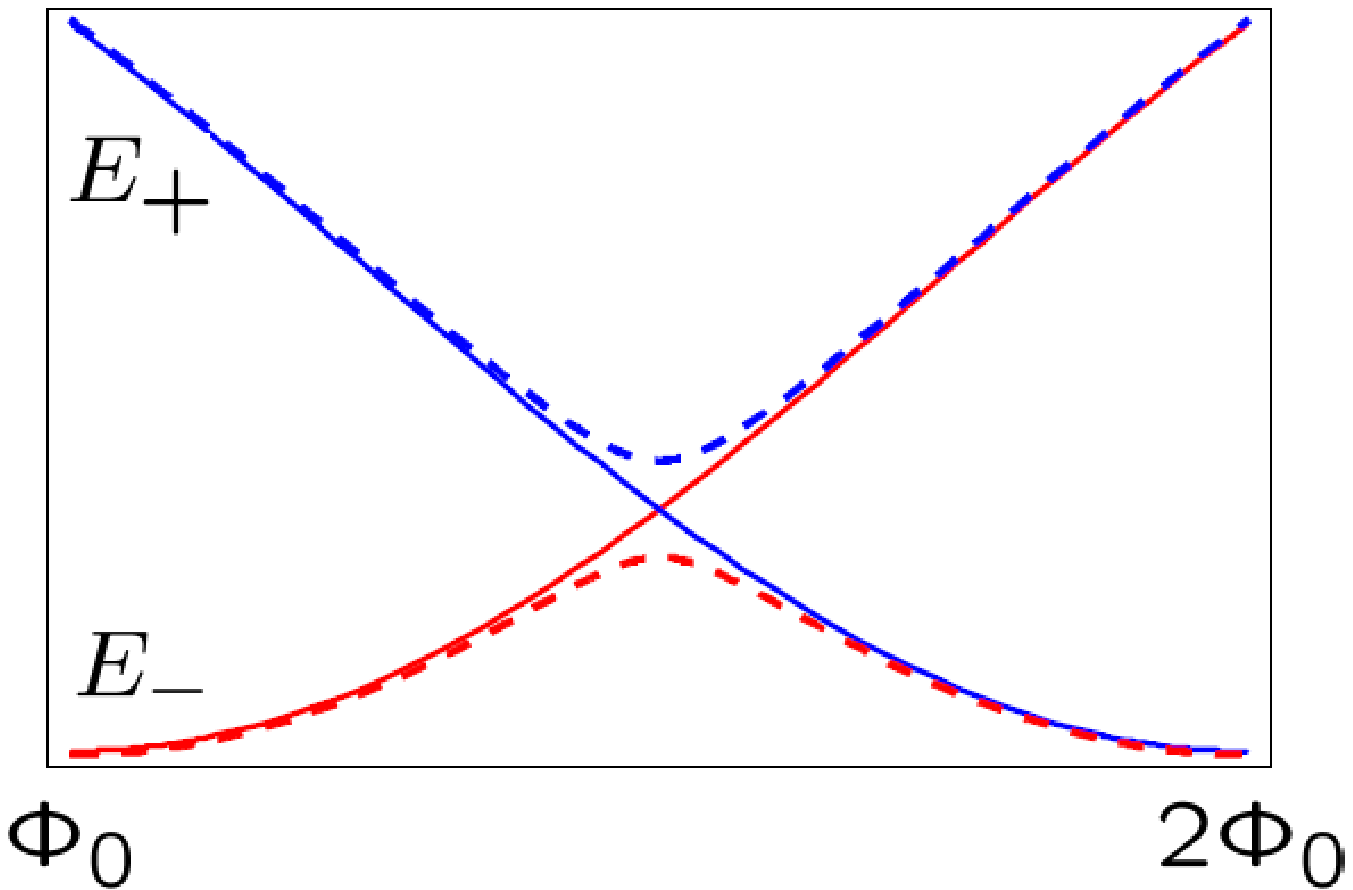}
\caption{(Color online) Left: Evolution of energy levels of a TQD in
a magnetic field. Right: Zoomed level crossing around
$\Phi=3\Phi_0/2$.}\label{f.4}
\end{figure}

The triangular geometry provides us with a new possibility for
controlling quantum tunneling\cite{KKA}. The tunnel current may be
driven by means of an external magnetic field oriented normally to
the plane of triangle, because the electron spectrum of electrons
in the TQD is a function of the magnetic flux $\Phi$ through the triangle
in such a geometry. As a result orbital degrees of freedom may be
activated at finite magnetic field, and the possibility opens to
realize a charge-spin conversion mechanism already in the case of
the TQD occupied by a single electron (${\cal N}=1$). The electron
energy spectrum at finite $\Phi$ is
\begin{equation}\label{emag}
\varepsilon_{\Gamma}(\Phi)=\epsilon-2V\cos
\left(p-\frac{\Phi}{3}\right).
\end{equation}
 and the values
of $p=0,~2\pi/3,~4\pi/3$ correspond respectively to $\Gamma=A,E_+,
E_-$. At zero $\Phi$ the states $\varepsilon_{E\pm}$ form degenerate
doublet excitations. At finite $\Phi$ the electron acquires chirality,
the doublet is split, the levels evolve in accordance with
(\ref{emag}) and at $\Phi_n=(n+ 1/2)\Phi_0$ the ground state becomes
doubly degenerate (here $\Phi_0$ is a quantum of magnetic flux,
$n=\pm1,\pm2, \ldots$). The level evolution is shown in Fig.
\ref{f.4}a.
The dynamical symmetry $SU(4)$ is involved in the signal
transformation at $\Phi$ close to $\Phi_n$, or in other words, at
$\varepsilon_\Gamma - \varepsilon_{\Gamma'}\sim T_K$ for any pair
of states $\Gamma,\Gamma'$. Figure \ref{f.4}a shows the evolution
of the energy levels in the interval $0 \leqslant \Phi \leqslant
3\Phi_0$, but the real periodicity of the spectrum is of course
$\Phi_0$ because at finite magnetic field all three states may be
converted into each other by the appropriate choice of the
 gauge. To make the notation symmetric we will consider below
 the sector $\Phi_0 \leqslant \Phi\leqslant 2\Phi_0$ shown in
 Figure \ref{f.4}(b). In this
 sector two states $E_{\pm}$ become nearly degenerate at $\Phi$
 approaching $3\Phi_0/2$. When the symmetry of TQD is perfect, the
 point $\Phi=3\Phi_0/2$ is a point of accidental degeneracy, in
 which the symmetry of TQD is $SU(4)$ \cite{KKA}. This
 symmetry is broken by the potential $v_{g}(t)$. As a result level
 crossing transforms into anticrossing (dashed lines in Fig.
 \ref{f.4}, left panel).

 In order to calculate the contribution of $\delta H_{\rm
dot}(t)$ in the charge-spin conversion we need the matrix elements of
the operators $T_{1j}^{(\pm)}$ (\ref{thop}) in the basis
$|\Gamma\rangle$ of the $C_{3v}$ point group,  in particular those
involved in the above level crossing/anticrossing in the sector
$\Phi_0 \leqslant \Phi\leqslant 2\Phi_0$. The relevant matrix
elements in the subspace $|E_{\pm}\rangle$ are
\begin{eqnarray}\label{tmel}
&&\langle E_\pm |H_{dot}^{(1)}|E_{\pm}\rangle= -2V\sin\varphi
\phi_1(t), \nonumber \\
&&\langle E_\pm
|H_{dot}^{(1)}|E_{\mp}\rangle=0
\end{eqnarray}
and
\begin{eqnarray}\label{tme2}
&&\langle E_\pm |H_{dot}^{(2)}|E_{\pm}\rangle=- V\cos\varphi\,
\phi^2_1(t),\nonumber \\
&&\langle E_\pm
|H_{dot}^{(2)}|E_{\mp}\rangle=-Ve^{i\varphi}\phi^2_1(t)
\end{eqnarray}
It follows from (\ref{tmel}), that the first-order term gives a
purely adiabatic contribution $\sim  \tilde\phi_1(t)$
(\ref{phi2}). Second-order corrections (first line in Eq.
\ref{tme2}) slightly change the adiabatic renormalization.
The stochastic signal $\sim \overline{\phi_1^2(t)}$ arises from the
off-diagonal matrix elements (second line in Eq. \ref{tme2}).

\subsection{Coherent input signal}

Following Ref. \onlinecite{KKA}, we use the irreducible
representations $\{A,E_\pm\}$ not only for dot eigenstates
(\ref{A1}) but also for lead states
\begin{eqnarray}\label{A2}
    &&c^\dag_{A,\sigma}=\sum_k
(c^\dag_{1{k}\sigma}+c^\dag_{2k\sigma}+c^\dag_{3k\sigma})/\sqrt{3}~,\\
&&c^\dag_{E_\pm,\sigma}= \sum_k(c^\dag_{1k\sigma}+e^{\pm
2i\varphi}c^\dag_{2k\sigma}
    +e^{\pm i\varphi}d^\dag_{3k\sigma})/\sqrt{3}~. \nonumber
\end{eqnarray}
in the three-terminal geometry.

\begin{figure}[h]
\includegraphics[height=2.75cm,width=4.2cm,angle=0]{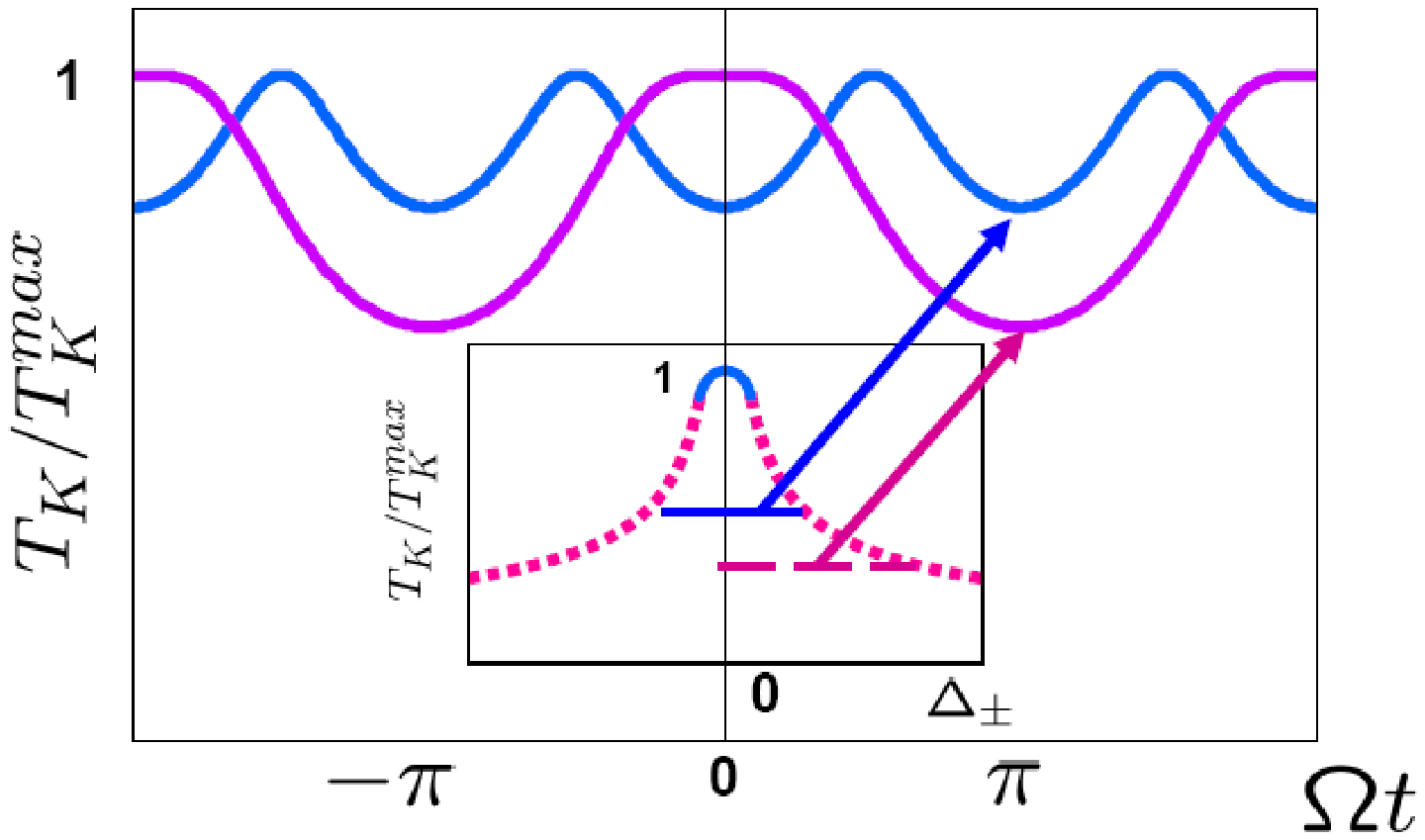}
\includegraphics[height=2.75cm,width=4.2cm,angle=0]{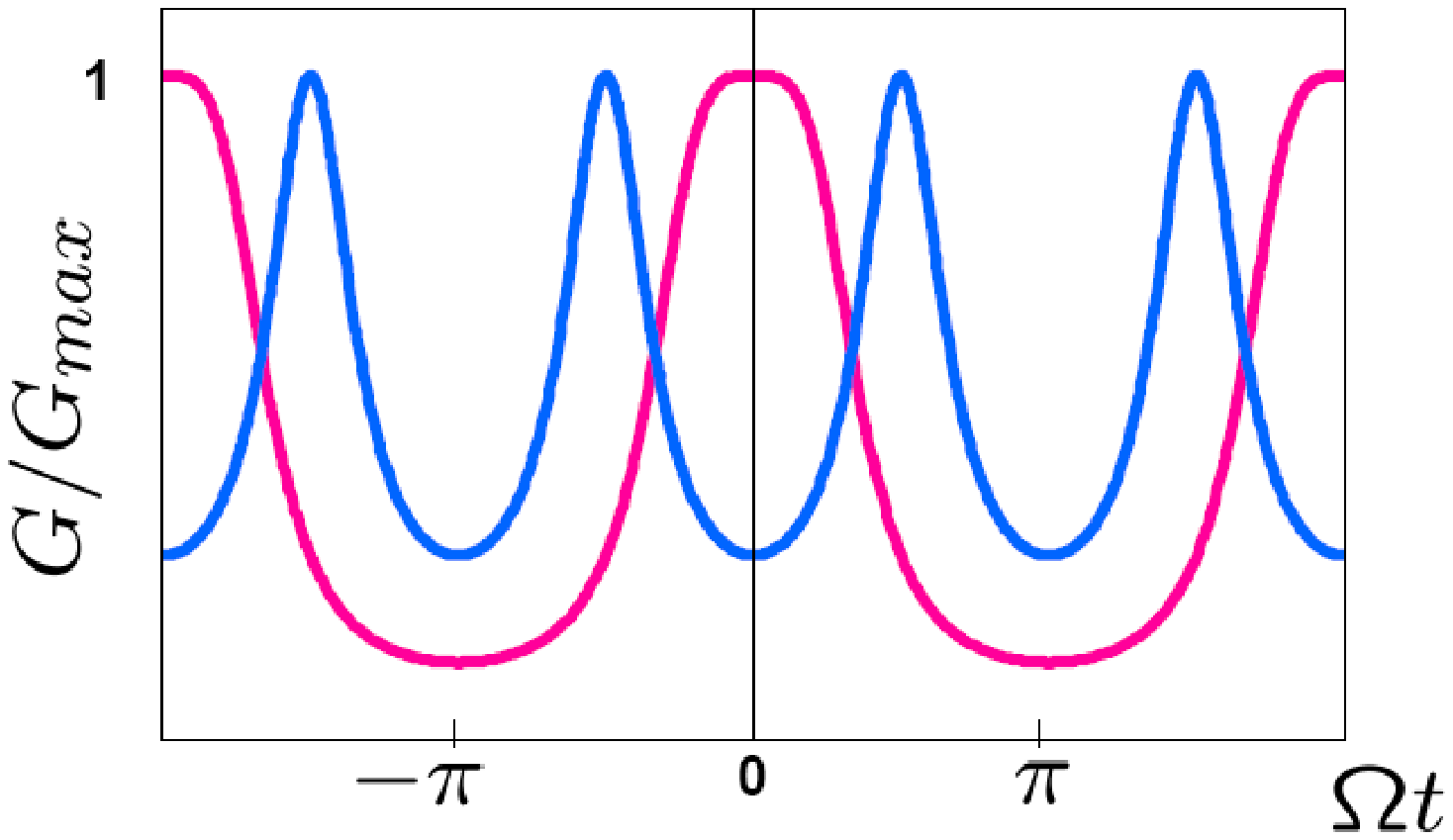}
\caption{(Color online) Left panel: Time dependent $T_K$
corresponding to the evolution of $\Delta_{\pm}$ in TQD. Inset: $T_K$
as a function of $\Delta_{\pm}$. The intervals of this evolution
are shown by straight lines. Right panel: Time dependent ZBA in the
conductance in accordance with the evolution of $T_K$.}\label{f.4aa}
\end{figure}

In zero magnetic field the ground state is degenerate, the singly
occupied TQD works as an effective spin 1/2, so that the effective
spin Hamiltonian has the standard form $J {\bf S}\cdot {\bf
s}_{AA}$, where only fully symmetric combinations with $\Gamma=A$
of lead and dot electrons are represented. Orbital degrees of
freedom become relevant near the crossing points, so that the
effective Kondo Hamiltonian has the form
\begin{eqnarray}\label{cotrst}
H_{cotun}= \sum_{\Gamma\Gamma'} J_{\Gamma\Gamma'} {\vec
S}_{\Gamma\Gamma'}\vec s_{\Gamma'\Gamma} + J_o \cal \vec T
{\vec\tau}
\end{eqnarray}
in the representation (\ref{A1}), (\ref{A2}). The spin operator for
lead electrons is determined as ${\bf
s}^i_{\Gamma\Gamma'}=\sum_{{\bf k}{\bf
k'}}c^\dag_{\Gamma,{\bf{k}}\sigma}\hat \tau_i c_{\Gamma',{\bf
k'}\sigma'}$. Due to the orbital degeneracy ($E_+ = E_- $), one
more vector, namely the pseudospin vector ${\cal\vec T}$ defined
as
\begin{eqnarray}
{\cal T}^{+}&=&\sum_\sigma |E_+,\sigma\rangle\langle
E_-,\sigma|, \ \ {\cal T}^{-}=[{\cal T}^{+}]^{\dagger},\\
{\cal T}^z&=&\frac{1}{2}\sum_\sigma\left(
|E_+,\sigma\rangle\langle E_+,\sigma|-|E_-,\sigma\rangle\langle
E_-,\sigma| \right).\nonumber
\end{eqnarray}
together with its counterpart for lead electrons is involved in
Kondo tunneling. There five vectors provide 15 generators of the
SU(4) group. The effective Kondo Hamiltonian consists of 6 terms
with corresponding exchange vertices, three of these vertices are
relevant (including that for pseudospin interaction), and the
corresponding Kondo temperature $T_K^{(E)}$ for $\Delta_{\pm}=0$
exceeds the zero-field Kondo temperature $T_K^{(A)}$ by a factor
of five (see Ref. \onlinecite{KKA} for details). Similarly to the
case of ST degeneracy, deviation from the level-crossing point
results in a sharp decrease of $T_K$, although in this case the
peak is symmetric relative to the zero gap point (Inset in the
left panel of Fig. \ref{f.4aa}). The width of this peak may be
estimated as $\sim T_K^{(E)}$.

In the vicinity of the level crossing point the $E_\pm$ orbital
components of the spin operators in (\ref{cotrst}) are involved in
Kondo screening, provided the difference $\Delta_{\pm}=|E_+-E_-|\sim
T_K$. Like in the case of a DQD (\ref{asymp}), the Kondo temperature itself
depends on the level distance, $T_K=T_K(\Delta_{\pm})$. We work in
the adiabatic regime and incorporate  matrix elements (\ref{tmel}),
(\ref{tme2}) in the energy terms $\varepsilon_{\pm}$ (\ref{emag}).
We conclude from these equations and from Fig. \ref{f.4} that only
the non-diagonal matrix elements $\langle E_\pm
|H_{dot}^{(2)}|E_{\mp}\rangle$ are relevant. Mixing of two branches
$\sim \phi_1^2$ results in a time-dependent lifting of degeneracy of
the orbital doublet at zero magnetic field, $\Phi=0$, and at $\Phi
\approx 3\Phi_0/2$. In agreement with the general rule for the Kondo
effect in presence of dynamical symmetry\cite{KKA,Eto,AK08},
$T_K(\Delta_{\pm})$ is maximum for $\Delta_{\pm}=0$, but unlike the
case of DQD (Fig. \ref{f.2}), the curve $T_K(\Delta_{\pm})$ is
symmetric around its maximum
(see Fig. \ref{f.4aa}, inset to the right panel). Now we may repeat
the procedure proposed for the DQD in the previous section and calculate
the time-dependent response of the ZBA peak in tunnel conductance to a
coherent input signal (\ref{tme2}). Two types of adiabatic temporal
oscillations $T_K(t)$ are presented in the left panel of Fig.
\ref{f.4aa}. If $\Delta_\pm$ oscillates between zero and some
maxima due to fluctuations of the gate voltage, $T_K(t)$ reaches
its maximum at $\Omega t = 2\pi n$. The tunnel conductance $G(t)$
shown in the right panel evolves with the same periodicity in
accordance with Eq. (\ref{conduct}). If $\Delta_\pm$ oscillates
symmetrically around zero value, the period is halved, and $G(t)$
reaches minima at $\Omega t = \pi n$. In the general case when
$\Delta_\pm$ varies between $-\Delta_a$ and $+\Delta_b$, the
oscillations of $G(t)$ are periodic but not monochromatic.
Experimentally one may turn from one regime to another by changing the
magnetic field (shifting the value of $\Phi$) in the vicinity of the
point $\Phi=3\Phi_0/2$.

We see that the situation with the $SU(2) \to SU(4)\to SU(2)$
crossover is close to the case of the $SO(4)\to SO(5)\to U(1)$
crossover discussed in the previous section from the point of view
of the conversion of the charge signal into Kondo response.  Due to the
fact that the orbital degrees of freedom are involved in the
formation of an effective exchange, the perturbation $v_g(t)$
directly affects Kondo tunneling by means of a time-dependent
lowering of the point symmetry of the triangle induced by the gate
voltage.

\subsection{Incoherent input signal}

The stochastization of orbital degrees of freedom in a TQD is induced by
the term $\delta H^{(\rm stoch)}_{dot}(t)$ (\ref{htime}),
(\ref{tme2}). The relevant part is
\begin{equation}
\Delta^{(\rm stoch)}(t)=\langle E_-|H_{\rm dot}^{\rm
stoch}|E_+\rangle = -V\overline{\phi_1^2(t)}e^{i\varphi}
\end{equation}
These stochastic inter-level transitions are responsible for the
{\it fluctuation induced} avoided crossing  $E_+  - E_-$ (dashed
lines in the right panel of Fig. \ref{f.4}). One may write the
corresponding part of the Hamiltonian in terms of pseudospin
operator $\cal \vec T$, as
\begin{equation}\label{trst}
H_{\rm dot}^{\rm stoch}= \Delta_\pm^{(\rm stoch)}(t) {\cal T}^- +
\Delta_\pm^{*(\rm stoch)}(t) {\cal T}^+
\end{equation}
This term should be added to the effective Kondo Hamiltonian of
TQD (\ref{cotrst}).

Thus in this case the TQD stochastic fluctuations of gate voltage
induce random pseudospin scattering, and the Keldysh approximation
(\ref{fluct}) may be used for this type of random potential.
However, unlike the DQD case, the scattering has a vector character,
so that the fermionized Hamiltonian (\ref{trst}) has the form
\begin{equation}\label{hfermt}
H^{\rm stoch}_{\rm dot}(t)= \varrho(t) g^{\dag}_\downarrow
g^{}_\uparrow + \varrho^*(t)g^{\dag}_\uparrow g^{}_\downarrow
\end{equation}
where $\varrho(t)$ is a random scattering potential which stems from
(\ref{trst}),   $g_\uparrow$ and $g_\downarrow$ are
"pseudospin-fermions" for vector $\cal \vec T$. Only the transversal
component of pseudospin scattering is involved in the stochastic
perturbation.

Then, following the pattern of the scalar model (\ref{fluct}), we
introduce the correlation function $C(t-t')=\langle
\varrho(t)\varrho^*(t')\rangle$ and its Fourier transform $\sim$
Gaussian variance $\xi$,
\begin{equation}
C(\omega)=\lim_{\gamma\to 0}
\frac{4\xi^2\gamma}{\omega^2+\gamma^2}=4\pi \xi^2\delta(\omega)
\label{fluctvec}
\end{equation}

The expansion of the Fourier transform of the Green's function for
pseudospin operators
\begin{equation}\label{fet}
F^R_{\sigma}(t-t')=\langle
g^{}_\sigma(t)g^\dag_\sigma(t')\rangle_R
\end{equation}
[cf. Eq. (\ref{get})] has the same form as in the scalar Keldysh
model, \cite{keld65,sad}, namely,
\begin{equation}\label{seriest}
F(\varepsilon) = f(\varepsilon) + \sum_{n=1}^\infty B_n
(\sqrt{2}\xi)^{2n}f^{2n+1}(\varepsilon).
\end{equation}
Here $f(\epsilon)=(\epsilon+i\delta)^{-1}$ is the free
spin-fermion propagator, $B_n$ is the total number of 2n-th order
diagrams. The indices $R,\sigma$ are omitted here and below for
the sake of brevity. The main advantage of the Keldysh model,
namely the equivalence of all diagrams corresponding to various
combinations of noise correlation functions in the self energy
(Fig. \ref{f.self1}) is still available. However, the essential part
of the diagrams in $\Sigma(\epsilon)$ disappears due to the kinematic
constraint ${\cal T}^+{\cal T}^+=0$ (or $g^\dag_\sigma
g^\dag_\sigma=0$). This means that only the diagrams with the
pseudospin operators  ordered as $\ldots{\cal T}^+{\cal T}^-{\cal
T}^+{\cal T}^-\ldots$ survive in the self energy (Fig.
\ref{f.self1}a).

A similar version of the cross technique in a real space arises in
electron systems in the domain of long-range Gaussian fluctuations
near the charge density wave (CDW) instability, although the
physical mechanism is radically different (alternating incoming and
outgoing Umklapp fluctuations of CDW order parameter in
1D\cite{sad,sad2} and 2D\cite{Barkop,Kusad} systems).\cite{footjetp}

\begin{figure}[h]
\includegraphics[width=7.0cm,angle=0]{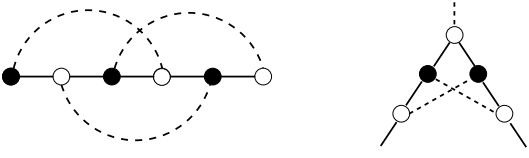}
\caption{First non-vanishing diagrams for the vertex (right) and
the vertex correction to the fermionic self energy (left) in the
vector Keldysh model. Black and white sites correspond to the two
terms in the Hamiltonian (\ref{hfermt}); pseudospin correlation
functions (\ref{fet}) are represented by dashed lines.}\label{f.5}
\end{figure}

One may represent the diagrams for the vector Keldysh model in the
following way. The vertices on the fermion line have two colors
(say, black and white) corresponding to the first and second term
in the time-dependent Hamiltonian (\ref{hfermt}). The black and
white vertices alternate and the wavy lines connect only the
vertices of different colors. Therefore the perturbation series
includes only the even order terms with equal number of black and
white vertices. Following these rules the vertex
correction presented by the 4-th order diagram of Fig.
\ref{f.self1}d disappears. One of nonzero 6-th order diagrams is
shown in Fig. \ref{f.5} (left). This is the first non-vanishing
vertex correction to the self energy of the vector Keldysh model.
 The difference between scalar and vector Keldysh model
is in this combinatorial coefficient.  In the scalar model the
coefficient $A_n=(2n-1)!!$, and summation of the perturbation
series results in Eq. (\ref{GR}). In the vector model the
coefficient $B_n=n!$ due to above kinematic selection rules, and
summation gives
\begin{widetext}
\begin{equation}\label{fvect}
F(\epsilon)=\frac{1}{2}\int_{-\infty}^{+\infty}
\frac{dxe^{-x^2/2\zeta^2}}{\zeta\sqrt{2\pi}}
\int_{-\infty}^{+\infty}
\frac{dye^{-y^2/2\zeta^2}}{\zeta\sqrt{2\pi}}
    \left[
\frac{1}{\epsilon-\sqrt{x^2+y^2}+i\delta}+\frac{1}{\epsilon+\sqrt{x^2+y^2
+i\delta}}
    \right].
\end{equation}
\end{widetext}
(see Appendix B for the derivation of this result). We see that Eq.
(\ref{fvect}) is the natural generalization of the one-dimensional
Gaussian averaging (\ref{GR}) characteristic for the scalar
Keldysh model to the two-dimensional Gaussian averaging of a vector
random field with purely transversal $xy$ fluctuations. Only the
modulus of the random field $r=\sqrt{x^2+y^2}$ is averaged with the
Rayleigh distribution function ${\cal P}_R(\epsilon)=
(\epsilon/\zeta^2)\exp(-\epsilon^2/2\zeta^2)$, whereas the angular
variable remains irrelevant due to the in-plane isotropy of the
system. Like in the scalar model, the averaged pseudospin-fermion
Green's function has no singularities.

In order to find the Ward identity for the vertex and the differential
equation for the Green's function generalizing Eqs. (\ref{vert}),
(\ref{DE}) we propose for the vector Keldysh model the procedure
which is an alternative to that used in \cite{efros70} for the scalar
model. We calculate explicitly the derivative
$dG^R_{\sigma}(\epsilon)/d\epsilon$ using the same method of
summation of the series in the r.h.s. of Eq. (\ref{seriest}) as the one
which was used for the calculation of the integral representation
(\ref{fvect}) for the Green's function. This calculation (see
Appendix B) gives the following result:
\begin{equation}\label{difvect}
\xi^2 \frac{dF(\varepsilon)}{d\epsilon} =1-\epsilon
F(\epsilon)\left(1-\frac{\xi^2}{\epsilon^2}\right)
\end{equation}
which is obviously the generalization of Eq. (\ref{DE}) with the
similar boundary condition $F(\epsilon\to\infty) =\epsilon^{-1}$.
One may rewrite Eqs. (\ref{DE}) and (\ref{difvect}) in a unified
way:
\begin{eqnarray}\label{dif}
\epsilon G(\epsilon) -1 &=&
\zeta^2G^2(\epsilon)\frac{d}{d\epsilon}G^{-1}(\epsilon) \nonumber \\
\\
\epsilon F(\epsilon)-1 &=& \xi^2
F^2(\epsilon)\left\{\frac{1}{\epsilon}\frac{d}{d\epsilon}\left[
\epsilon F^{-1}(\epsilon)\right] \right\}\nonumber
\end{eqnarray}

One may check straightforwardly that the functions $G(\epsilon)$
and $F(\epsilon)$, given by Eqs. (\ref{GR}) and (\ref{a2.int2})
are indeed solutions of differential equations (\ref{dif}). Then,
appealing to Eq. (\ref{vert}), we define the vertex
$\Gamma(\epsilon,\epsilon,0)$ for the vector model as
\begin{equation}\label{ward}
      \Gamma = \frac{1}{\epsilon}\frac{d}{d\epsilon}
   (\epsilon
   F^{-1}(\epsilon)).
\end{equation}
It is worth noting that the differential operator in the r. h. s.
of Eq. (\ref{ward}) is nothing but ${\rm div}_\epsilon$ in polar
coordinates. This form reflects effective two-dimensionality of
Gaussian averaging in the vector Keldysh model, which has been
noticed already in Eq. (\ref{fvect}). Equations (\ref{difvect})
and (\ref{ward}) facilitate the calculation of response functions
of TQD.

The stochastization of pseudospin manifests itself in the
transformation of the response function
$\chi_\perp(t)=\langle{\cal T}^-(t){\cal T}^+(0)\rangle_R$ shown
in Fig. \ref{f.loop1}. Now the solid lines correspond to
pseudofermion propagators $F_\uparrow(t)$ and $F_\downarrow(-t)$,
and the vertex corrections are presented by the diagram \ref{f.5}
(right panel) and the higher-order diagrams of that sort. In order
to calculate the pseudospin susceptibility at finite temperature we
address to the equation
\begin{eqnarray}\label{chiper}
&&\chi(i\omega_m)=\\
&& T\sum_n{\cal F}_\uparrow(i\omega_m+i\epsilon_n){\cal
F}_\downarrow(i\epsilon_n)
\Gamma(i\epsilon_n,i\epsilon_n+i\omega_m;i\omega_m)\nonumber
\end{eqnarray}
similar to Eq. (\ref{chi}). Here ${\cal F}_\sigma(i\epsilon_n)$ is
the Matsubara-type analytical continuation of the Fourier
transform of the Green's function (\ref{fet}). Using the definition
(\ref{ward}) of the vertex $\Gamma$ and the second equation from
(\ref{dif}), we express the vertex as
\begin{equation}
\Gamma(\epsilon_n,\epsilon_n,0)= \frac{i\epsilon_n {\cal
F}(i\epsilon_n)-1}{\xi^2 {\cal F}^2(i\epsilon_n)}
\end{equation}
[cf. Eq. (\ref{GamR})]. Following the procedure which led to Eqs.
(\ref{ssuca}) and (\ref{Gm}) in scalar model, we obtain for
$\chi_\perp(0)$ and its asymptotics the following equations
\begin{eqnarray}\label{ssucv}
\chi_\perp(0)=\frac{1}{\xi}\int_{0}^{\infty}y^2dy e^{-y^2/2
}\tanh\left(\frac{y\xi}{2T}\right)
\end{eqnarray}
and
\begin{eqnarray}
\chi(0)\sim \left\{
\begin{array}{c}
1/T,\;\;\; T\gg \xi,\\
    1/\xi,\;\;\; \xi\gg T \label{Gmb}
\end{array}\right.
\end{eqnarray}
Thus the pseudospin in the vector model looses its local
characteristics in the same way as the spin in the scalar model.
Accordingly, stochastization affects the Kondo tunneling. To
estimate this effect we calculated the electron-pseudofermion loops
similar to those shown in Fig. \ref{f.loop2}, but with solid lines
standing for ${\cal F}_\sigma(i\epsilon_n)$. Since the weak coupling
approach works only in the limit $T,\xi \gg \Delta_\pm$, the
pseudo-gap in the spectrum does not affect the logarithmic behaviour
of the Kondo-loop. The stochastization-induced level repulsion in
the vector model excludes states within the pseudogap both when
$T\gg\xi$ and $T\ll\xi$. In the latter case the infra-red  Kondo
cutoff is of the order of $\xi$. Therefore, the resonance Kondo
tunneling in this case is controlled by spin degrees of freedom only
in accordance with the $SU(2)$ Kondo effect paradigm.

We conclude from the above results, that the case of $SU(4)$ symmetry
supported by the interplay between spin and orbital degrees of freedom
in a TQD differs radically from the case of $SO(4)$ symmetry
involving only the spin variables. In the latter case the external
charge noise results in the stochastization of spin degrees of
freedom, so that the DQD "looses" its spin moment at low enough
energy/temperature. In the former case the spin 1/2 is robust, and
only the orbital (pseudospin) degrees of freedom are affected by
the charge noise. Pseudospin stochastization means that the
logarithmic divergences in the corresponding Kondo loops given by the
diagrams similar to those of Fig. \ref{f.loop2} are subject to a
cut-off similar to that given by Eq. (\ref{Gm}). As a result, only
the spin-electron loops determine the Kondo screening at low $T$.
Thus a {\it noise induced $SU(4)\to SU(2)$ crossover takes place
in TQD}.

\section{Concluding remarks}

In this paper we have demonstrated that the charge-to spin
conversion of a time dependent input signal applied to the gate
attached to a side well of a complex quantum dot is possible through
several mechanisms involving the dynamical symmetry of this
nano-object. Such a possibility arises when the multiplet
 involved in the dynamics of low-energy excitations includes {\it
 both}
 charge and spin degrees of freedom. In the case of $SO(5)$ symmetry
 characterizing the T-shaped DQD with even occupation, charge
 transfer singlet excitons are activated by $v_g(t)$, and the spin
 degrees of freedom are excited via the Casimir constraint. In the
 case of $SU(4)$ symmetry which determines the dynamics of TQD with
 single electron occupation, the signal $v_g(t)$ affects orbital
 (pseudospin) degrees of freedom, which are involved in the Kondo
 screening together with conventional spin states.

 Both coherent and stochastic components of $v_g(t)$ are subject
 to a charge-spin transformation. The Kondo response to a coherent
 charge signal is close in its nature to oscillations of the Kondo
 temperature discussed in a context of Kondo shuttling.\cite{shut}
The noise component introduces the stochastization of spin degrees
of freedom. This stochastization may be complete in DQD, provided
the variance of the Gaussian noise exceeds $T_K$. Then the low-energy
cutoff results in the smearing and even the elimination of the
Kondo-related ZBA in tunneling conductance. In TQD with $SU(4)$
symmetry only the charge (orbital) degrees of freedom are
stochasticized. Strong enough noise may result in peculiar
noise-induced quantum crossover $SU(4)\to SU(2)$, which may be
controlled experimentally by varying the noise level in the input
signal.

\appendix

\section{}

The time-dependent SW transformation may be performed in the
adiabatic approximation.\cite{KNG} In our case it gives the following
expressions for the coupling constants controlling $T/S$ and $T/E$
transitions:\cite{KKAR}
\begin{equation}\label{2.3a}
J^{S}(t)=\frac{W-w_2^{S}(t)}{\epsilon_2-M_S(t)}
\cdot\frac{W(1-V/\Delta_{ES}(t))\sqrt{2}\phi_1(t)}{\Delta_{ES}(t)},
\end{equation}
\begin{equation}\label{2.3b}
J^{E}(t)=\frac{W-w_2^{E}(t)}{\epsilon_2 +Q_2+M_E(t)}
\cdot\frac{W(1-V/\Delta_{ES}(t))\sqrt{2}\phi_1(t)}{\Delta_{ES}(t)},
\end{equation}
with
$$
M_S(t)= M_S - C_S\phi_1^2(t),\;\;\; M_E(t)= M_E + C_S\phi_1^2(t)
$$\\
$$
w_2^{S}(t)=\frac{VW\sqrt{2}\phi_2^{S}(t)}{\Delta_{ES}},\;\;\;
w_2^{E}(t)=\frac{VW\sqrt{2}\phi_2^{E}(t)}{\Delta_{ES}},
$$

$$
\phi_2^{S}(t)=\phi_1(t)-\kappa_2^{S}(t),\;\;\;
\phi_2^{E}(t)=\phi_1(t)-\kappa_2^{E}(t),
$$
$$
\kappa_2^{S}(t)= v_g(t)/\epsilon_2,\;\;\; \kappa_2^{E}(t)=
v_g(t)/(\epsilon_2  +Q_2).
$$

To second order in $v_{g}(t)$ Eqs. (\ref{2.3a}) and (\ref{2.3b})
lead to the averaged time-dependent corrections to the coupling
coefficients
\begin{eqnarray}\label{Apex}
&&\overline{J^{S}(t)} = J^S_{\rm ad} +  J^S_{\rm stoch} \\
&& \approx \frac{\sqrt{2}W^{2}}{\epsilon_2-M_S}
\left(\tilde{\phi}_1(t) + \frac{\sqrt{2}V}{\Delta_{ES}}
\left(\frac{\overline{v_{g}(t)\phi_1(t)}}{\epsilon_1} -
\overline{\phi_1(t)^{2}}\right)\right) \nonumber
\end{eqnarray}
and a similar expression for $\overline{J^{E}(t)}$ where
$\epsilon_2-M_S$ is replaced by $\epsilon_2+Q_2 + M_E$.

\section{}

To derive the pseudospin-fermion GF for the vector Keldysh model,
we generalize the original Keldysh summation
procedure\cite{keld65,sad} to the perturbation series
(\ref{seriest}). Using the integral representation for the
$\Gamma$-function, $ n!=\int_0^\infty dz z^ne^{-z}, $ we transform
the series (\ref{seriest}) into
\begin{eqnarray}
F(\epsilon)=f(\epsilon)\left\{
    1+2\sum_n\int_0^\infty tdt
    \left[t\xi f(\epsilon)]^{2n}
    \right]e^{-t^2}
\right\}\nonumber
\end{eqnarray}
(cf. Ref. \onlinecite{sad2}). Here we substituted $t^2$ for the
variable $z$. Then changing the order of summation and
integration, we transform $F_\sigma(\epsilon)$ into the integral
\begin{equation}\label{a2.int1}
F(\epsilon)= \int_0^\infty
    2tdt\frac{f(\epsilon)}{1-2t^2\xi^2f^2(\epsilon)}e^{-t^2}
\end{equation}
Taking into account the explicit form of the free
pseudospin-fermion propagator,
$f\epsilon)=(\epsilon+i\delta)^{-1}$, we  change the integration
variable once more, $t=u/\sqrt{2}\xi$ and put it into the form
(\ref{a2.int1})

\begin{equation}\label{a2.int2}
F(\epsilon)=\int_0^\infty \frac{udu}{2\xi^2}\left(
    \frac{1}{\epsilon -u+i\delta}+ \frac{1}{\epsilon+u+i\delta}\right)
    e^{-u^2/2\xi^2}
\end{equation}
Next we introduce the "cartesian" coordinates, $x=u\cos \phi,~
y=\sin\phi$, so that $u=\sqrt{x^2+y^2}$ and $dxdy=udud\phi$. The
angle-independent integral (\ref{a2.int2}) may be rewritten as
\begin{eqnarray}\label{a2fvect}
&&F(\epsilon)=\frac{1}{4\pi\xi}\int_{-\infty}^{+\infty}
dx\int_{-\infty}^{+\infty} dy e^{-(x^2+y^2)/2\xi^2}\nonumber\\
&&    \left[ \frac{1}{\epsilon-\sqrt{x^2+y^2}+i\delta}+
\frac{1}{\epsilon+\sqrt{x^2+y^2}+i\delta}
    \right]
\end{eqnarray}
which is in fact the expression in Eq. (\ref{fvect}).

In order to calculate the derivative $dF^R_\sigma/d\varepsilon$, we start
with  the same expansion (\ref{seriest}). The analog of Eq.
(\ref{a2.int1}) for the derivative has the form
$$
\frac{dF(\varepsilon)}{d\varepsilon}=
-f^2(\varepsilon)\left[1+\int_0^\infty 2tdt
\frac{2(2t^2-1)t^2\xi^2
f^2(\varepsilon)}{1-2t^2\xi^2f^2(\varepsilon)} e^{-t^2}\right]
$$
The subsequent variable change which gave Eq. (\ref{a2.int2}) for
the GF gives for its derivative the following equation
\begin{equation}\label{deriv}
\frac{dF(\varepsilon)}{d\varepsilon}= -f^2(\varepsilon)\left[1+
\frac{1}{2}\left(\frac{J_4}{\xi^4}-\frac{J_2}{\xi^2}
\right)\right]
\end{equation}
where
$$
J_n =\int_0^\infty dz z^n
\exp\left(-\frac{z^2}{2\xi^2}\right)\left[f(\varepsilon-z)-
f(\varepsilon+z) \right]
$$
After some manipulations, these integrals are represented via the
GF for the vector model (\ref{a2.int2}):
\begin{equation}
J_2 = 2\varepsilon \xi^2 F -2\xi^2,~~ J_4= -4\xi^4+ \varepsilon^2
J_2
\end{equation}
Substituting these integrals in Eq. (\ref{deriv}), we come after
some algebra to the differential equation (\ref{difvect}).


\begin{thebibliography}{99}
\bibitem{Dyak71} M.I. Dyakonov and V.I. Perel, Sov. Phys. JETP
Lett. {\bf13}, 467 (1971).

\bibitem{Munaz} S. Murakami, N. Nagaosa and S.-C. Zhang, Science
{\bf301}, 1348 (2003); N. Sugimoto, S. Onoda, S. Murakami, and N.
Nagaosa, Phys. Rev. B {\bf73}, 113305 (2006).

\bibitem{Dyak07} M.I. Dyakonov, Phys. Rev. Lett. {\bf99}, 126601
(2007).

\bibitem{Lug} H.-F.L\"u and Y. Guo, Appl. Phys. Lett. {\bf91},
092128 (2007)

\bibitem{Cacca76} C. Caccamo, G. Pizzimenti, and M.P. Tosi, Nuovo
Cimento {B\bf31}, 53 (1976).

\bibitem{Davi01} I. D'Amico and G. Vignale, Phys. Rev. B {\bf62},
4853 (2000).

\bibitem{Li} S. Li, T.F. George, X. Sun, and L.-S. Chen, J.
Phys. Chem. B {\bf111}, 6099 (2007).

\bibitem{KKAR} M.N. Kiselev, K. Kikoin, Y. Avishai, and J.
Richert, Phys. Rev. B {\bf74}, 115306 (2006)

\bibitem{Nova} K. Kikoin, Y. Avishai and M.N. Kiselev,
Dynamical symmetries in nanophysics, {\it in} "Nanophysics,
Nanoclusters and Nanodevices,"  (Nova Publishers, New York, 2006)
pp. 39-86; arXiv:cond-mat/0407063.

\bibitem{KKA} T. Kuzmenko, K. Kikoin and Y. Avishai, Phys. Rev.
Lett. {\bf 96}, 046601 (2006).

\bibitem{amato08} A. Amato,  T. Hatano, T. Kubo, Y. Tokura, D.G.
Austing, and S. Tarucha. Physica E {\bf40}, 1322 (2008).

\bibitem{KNG} A. Kaminski, Yu. V. Nazarov and L.I. Glazman, Phys. Rev. B
{\bf 62}, 8154 (2000).

\bibitem{BRUS} C. Bruder and H. Schoeller, Phys. Rev. Lett. {\bf 72}, 1076 (1994).

\bibitem{KA01} K. Kikoin and Y. Avishai, Phys. Rev. Lett. {\bf86},
2090 (2001); Phys. Rev. B {\bf 65}, 115329 (2002).

\bibitem{shut} M.N. Kiselev, K. Kikoin, R.I. Shekhter, and V.M.
Vinokur, Phys. Rev. B {\bf 74}, 233403 (2006).

\bibitem{Zeh} H.D. Zeh in {\it Decoherence and the Appearance of a
Classical World in Quantum Theory}, eds. E. Joos et al
(Springer-Verlag, Berlin, 2003) 2-nd ed., Chapter 9.

\bibitem{Hald} F.D.M. Haldane, Phys. Rev. Lett. {\bf40}, 416
(1978).

\bibitem{Anders} P.W. Anderson, J. Phys. C {\bf 3}, 2436 (1970).

\bibitem{Kogan} A. Kogan, G. Granger, M.A. Kastner, and D.
Goldhaber-Gordon, H. Shtrikman, Phys. Rev. B {\bf67}, 113309 (2003).

\bibitem{Eto} M. Eto and Yu.V. Nazarov, Phys. Rev. Lett. {\bf85}, 1306 (2000),
Phys. Rev. B {\bf64}, 085322 (2001); M. Eto, J. Phys. Soc. Jpn,
{\bf74}, 95 (2005).

\bibitem{Pust1} M. Pustilnik and L.I. Glazman, Phys. Rev. Lett.
{\bf85}, 2993 (2000), Phys. Rev. B {\bf64}, 045328 (2001).



\bibitem{kork} F. Delgado, Y.-P. Shim, M. Korkusinski, P.
Hawrylak, Phys. Rev. B {\bf76}, 115332 (2007)

\bibitem{kisrew} M.N. Kiselev, Int. J. of Mod. Phys. B,  {\bf 20},
381 (2006).

\bibitem{Edw} S.F. Edwards, Phil. Mag. {\bf 3}, 33, 1020 (1958).

\bibitem{footkayan} Similar problem of stochastization of the
Landau-Zener (LZ) level crossing problem in two limiting cases
(\ref{whno}) and (\ref{fluct}) was considered by Y. Kayanuma, J.
Phys. Soc. Jpn. {\bf 53}, 108 (1984); {\bf 54}, 2037 (1985),
detailed theory of slow-noise-assisted LZ problem is given in Ho
Ngoc Phien and M.N.Kiselev (to be published).

\bibitem{keld65} L.V. Keldysh, "Semiconductors in strong electric
field", D. Sci. thesis, Lebedev Institute, Moscow, 1965.

\bibitem{efros70} A.L. Efros, Sov. Phys.--JETP {\bf32},
479 (1971) [Zh. Eksp. Teor. Fiz. {\bf59}, 880 (1970)].

\bibitem{sad} M.V. Sadovskii, {\it Diagrammatics} (World Scientific, Singapore,
2005).

\bibitem{sad2} M.V. Sadovskii, Zh. Eksp. Teor.
Fiz. {\bf 66}, 1720 (1974) [Sov. Phys. -- JETP {\bf 39}, 845
(1974)].

\bibitem{Barkop} L. Bartosch and P. Kopietz, Eur. Phys. J. B
{\bf17}, 555 (2000).

\bibitem{Kusad} E.Z. Kuchinskii and M.V. Sadovskii, JETP {\bf94},
654 (2002).

\bibitem{footjetp} General formulation of Keldysh model in time
domain with both scalar and vector fluctuations taken into
account, was developed for the toy model of ensemble of two-level
systems in Ref. M.N. Kiselev and K. Kikoin, JETP Letters {\bf89},
114 (2009) [Pis'ma v ZhETF, {\bf89}, 133 (2009) ].

\bibitem{footkeld} The derivation procedure is outlined in
Appendix B in concern to a generalized (vector) Keldysh model.

\bibitem{kisop00} M.N. Kiselev and R. Oppermann, JETP Lett {\bf 71},
250 (2000) [Pis'ma v ZhETF {\bf 71}, 359 (2000)].

\bibitem{AK08} Y. Avishai and K. Kikoin,
Tunneling through Quantum Dots with Discrete Symmetries, in {\it
"Encyclopedia of Complexity"} (Springer Verlag, Berlin, 2008) ;
arXiv:0801.3095

\end{thebibliography}
\end{document}